\documentclass[pra,aps,amsmath,amssymb,twocolumn,floatfix,superscriptaddress]{revtex4-2}
\usepackage{amsmath}
\usepackage{graphicx}
\usepackage{color}
\usepackage{amssymb}
\usepackage[utf8]{inputenc}
\usepackage[T1]{fontenc}

\usepackage{comment}
\usepackage[normalem]{ulem}
\usepackage{hyperref}
\hypersetup{colorlinks,
 linkcolor=blue,%
 citecolor=blue,%
 urlcolor=blue
}

\usepackage{array}

\begin{document}
\title{Collective excitation spectra of dipolar bosonic fractional quantum Hall states }
\author{Moumita Indra \footnote{Orcid ID: 0000-0002-7900-7947}} 
\email{indram@iitg.ac.in}
\author{Pankaj Kumar Mishra \footnote{Orcid ID: 0000-0003-1566-0012}} 
\email{pankaj.mishra@iitg.ac.in}
\affiliation{Department of Physics, Indian Institute of Technology Guwahati, Guwahati - 781039, Assam, India}
\begin{abstract}
We numerically investigate the collective excitation of spin-conserving and spin-reversed  configuration of rotating diluted ultra-cold dipolar Bose gas. Rotating trapped Bose gas produces a fictitious magnetic field perpendicular to the trapping harmonic potential, which exhibits strongly correlated fractional quantum Hall states. We consider the long-range dipole-dipole interaction and compute the low lying  excitations spectrum for the three fractions of the first Jain series $\nu = 1/2, 1/4, 1/6$. We find that for both the spin-conserving and spin-reversing excitation the gap between the fundamental mode and the higher excitation mode increases upon increase in the filling fraction. The fundamental modes and the next higher-energy mode of excitation spectra for each of the three fractions show the presence of double roton for spin-conserving configuration only. Finally we complement our observation by calculating the spectral weight for the fundamental mode of excitation spectra which show the momenta at which the spectral weight exhibits the maxima shifts towards the lower momenta for both the excitations. Our observation for the spectral weight could be related with the inelastic Raman scattering which may be useful for the future experimental study to detect the excitation in ultracold system.
\end{abstract}
\maketitle
\section{Introduction}
Since its realization in the laboratory experiment, Bose-Einstein condensates (BECs)~\cite{Davis:1995, Anderson:1995} have provided an exceptional platform for studying collective excitations and quantum many-body phenomena. When interactions in a BEC are modified by long-range dipole-dipole interactions (DDIs)~\cite{Tang:2018}, the system exhibits rich and intriguing behavior that extends beyond conventional contact-interacting condensates~\cite{Regnault:2003}. Dipolar BECs, realized experimentally in ultra-cold gases of highly magnetic atoms such as Dysprosium and Erbium~\cite{Lu:2011, Trautmann:2018}, offer an unique avenue to explore anisotropic interactions, self-organized structures, and novel quantum phases. Following the experimental achievement of BEC, it became a central focus to study the elementary excitations~\cite{Yamamoto:2016}. In this context, Bogoliubov theory~\cite{Lieb:1963}, in conjunction with the Gross-Pitevskii equation~\cite{Rogel:2013}, provides a powerful framework for analyzing the dynamical properties and excitation spectrum of the condensate.


In recent years the studies on dipolar BECs have captured significant attention from the scientific community due to their immense potential application in the quantum technologies. The first dipolar BEC was realized in a gas of $^{52}$Cr~\cite{Griesmaier:2005} atoms and the nature of the dipolar interaction between atoms was regulated by a measurement of the expansion of condensate \cite{Stuhler:2005}.  $^{52}$Cr has magnetic dipole moment six times Bohr magnetos. After these discoveries, there are many studies in the field of dipole-dipole-interacting (DDI) BEC~\cite{Lahaye:2007, Olson:2013, Xi:2016}. Other dipolar species such as $^{162}$Dy, which has even larger magnetic dipole moment- about nine times the Bohr magneton- have also been condensed in the laboratory experiment~\cite{Lu:2011}.

Recent advances in the production and manipulation of BECs composed of highly magnetic atoms, particularly chromium and dysprosium, have fueled growing interest in long-range, anisotropic dipole-dipole interactions. These interactions play a crucial role in determining the static and dynamic characteristics of quantum degenerate states~\cite{Bismut:2010}. When such dipolar BECs are subject to rotation with trapping potentials, their collective excitation undergoes significant modifications, revealing new dynamical properties and potential instabilities~\cite{Lahaye:2009}. In particular, the interplay between dipolar interactions, trap geometry, and rotational effects leads to novel features in excitation modes, including roton-like minima~\cite{Chomaz:2018}, anisotropic phonon dispersion~\cite{Lyu:2022}, and formation  of vortex lattice structures~\cite{Hung:2012}. 

Rotating two-dimensional (2D) BEC system produces fictitious magnetic field that originates from the harmonic trap. This results the appearance of Landau-levels (LLs)~\cite{Greiter:2011} where, at high rotation and low density the bosons occupy the lowest levels. Under such formalism, the rotating trapped Bose atomic system~\cite{Chang:2005} known to exhibit the strongly correlated fractional quantum Hall effect (FQHE)~\cite{Tsui_PRL:1982, Tsui_PRB:1982, Stormer:1983, Prange:1990, Stormer:1992} that could not be explained using the conventional single-particle model~\cite{Struck:2006}. Using the classical plasma analogy, Laughlin first proposed the representation of  many-electron trial wave function in complex coordinates of an incompressible state~\cite{Laughlin:1983}. However theory could not accommodate the observation of all fractional states except the inverse of odd integers. There are some experiments that report the large choice for the  fractions that encompasses the even and odd denominators~\cite{Shi:2020, Chen:2024}. 

In last few decades work on the FQHE has seen significant rise in number due to its wide application in different areas. For instance Fractional quantum Hall (FQH) states can host quasi-particles with fractional charge and anionic statistics~\cite{Feldman:2021}, which allows robust topologically protected~\cite{Bruno:2012} nature of the materials such as quantized Hall conductance and Berry phase~\cite{Sprinkart:2024}.  Different FQH states could be modeled using the composite fermion (CF) theory first proposed by Jain~\cite{Jain:1990, Cooper:1999}. 

Under CF theory, the quasi-particles experience reduced effective magnetic field given by $B^{eff} = B - p \sigma \phi_0$, with $B$ as the fictitious magnetic field, $\phi_0=\frac{hc}{e}$ is the magnetic flux-quantum and $\sigma$ denotes the number density. The odd integer $p = 1, 3, 5, .. $ represents the number of flux quanta attached to the Bose atoms to form the quasi-particles. Those Bose composite fermions create a new type of LLs (called $\Lambda$-levels) in presence of weak effective magnetic field. The LL filling fraction of the Bose atoms ($\nu$) is related with the filling fraction of Bose-CF $\Lambda$-level  as $\nu = \frac{n}{n p + 1}$. Using this mapping, the composite particles can be treated like fermions~\cite{Leinaas:1977, Wilczek:1982} and that the filling fractions precisely match the Jain-series sequence~\cite{Jain:1989}. The filling fractions of Bose particles $\nu = 1/2, 1/4, 1/6$ maps with the one filled $\Lambda$-level ($n = 1$) with flux attachments $1, 3, 5$ respectively. Based on the fermionic Chern-Simons (CS) approach, Shankar and Murthy~\cite{Shankar:1997} developed a microscopic Hamiltonian theory of the FQHE~\cite{Lopez:1991}, which has been quite successful in explaining different fractions and calculating gaps in FQH states. This functional integral description performs different flux attachment via the CS gauge field~\cite{Modak:2011} to obtain for bosonic and fermionic particles.


Two-component BECs~\cite{Miesner:1999, Modugno:2002, Thalhammer:2008, Papp:2008} are generated by trapping magnetic hyperfine states of the same atomic species as well as mixtures of two distinct atomic species. Moreover, the two-component system raises the possibility of a more intricate FQH structure that includes partially spin-polarized states~\cite{Kukushkin:1999, Indra:2018}. The multi-flavored CF-picture (attaching different numbers of flux quanta with different species of electrons), $\Lambda$-levels within $\Lambda$-level has been studied to explain partially polarized Hall states beyond the Jain series~\cite{Balram:2015}. Rotating the trap that confines the atoms is a well-established approach to achieve quasi-degeneracy in atomic motional states. When the rotational frequency is low compared to the trap's frequency, the system forms a vortex condensate \cite{Abo:2001}, with the number of vortices increasing as the rotation speed increases \cite{Fetter:2009}. The vortex condensate is a weakly correlated state where all of the atoms can occupy same single-particle quantum state. However, as the rotation frequency approaches the harmonic trap frequency \cite{Baranov:2005}, the system is expected to enter the FQH regime, where the atoms become strongly correlated \cite{Pino:2013, Imran:2023}. 

While the collective excitations in rotating Bose gases especially in the context of quantum Hall fractions have been extensively studied~\cite{Das:2018, Indra:2018, Indra:2020, Indra:2023, Indra:2024}, the combined effects of rotation and dipolar interactions remain largely unexplored. Rotation modifies excitation spectra through vortex dynamics, while dipolar interactions introduce anisotropic roton-like modes. Understanding their interplay is the key to reveal new quantum phases which is the main aim our present study. In this paper, we have introduced the FQHE for a system of bosonic atoms confined in a very rapidly rotating trap, and examined collective spin-conserving and spin-reversed excitation of fully polarized FQH states of Bose atoms rather than different spin-polarization states.

The structure of our paper is as follows. In Section~\ref{sec:MF} we present the mean-field model used for our work. Further in Sec.~\ref{sec:waveFun} we discuss the geometry of the many-body wave function for the ground state as well as for the excited state for two types of collective excitations. Following this in the Sec.~\ref{sec:ExciSpec} we present excitation spectra of the dipolar rotating gas which followed by the analysis of the spectral weight in the Sec.~\ref{sec:SpecWeight}. Finally, in Sec.~\ref{sec:conclusion} we conclude our work and also provide a possible scenario the future realization of the finding presented in the current work.

\section{Mean-field model}\label{sec:MF}
\begin{figure*}[!htp]
\centering
\includegraphics[width=0.7\textwidth]{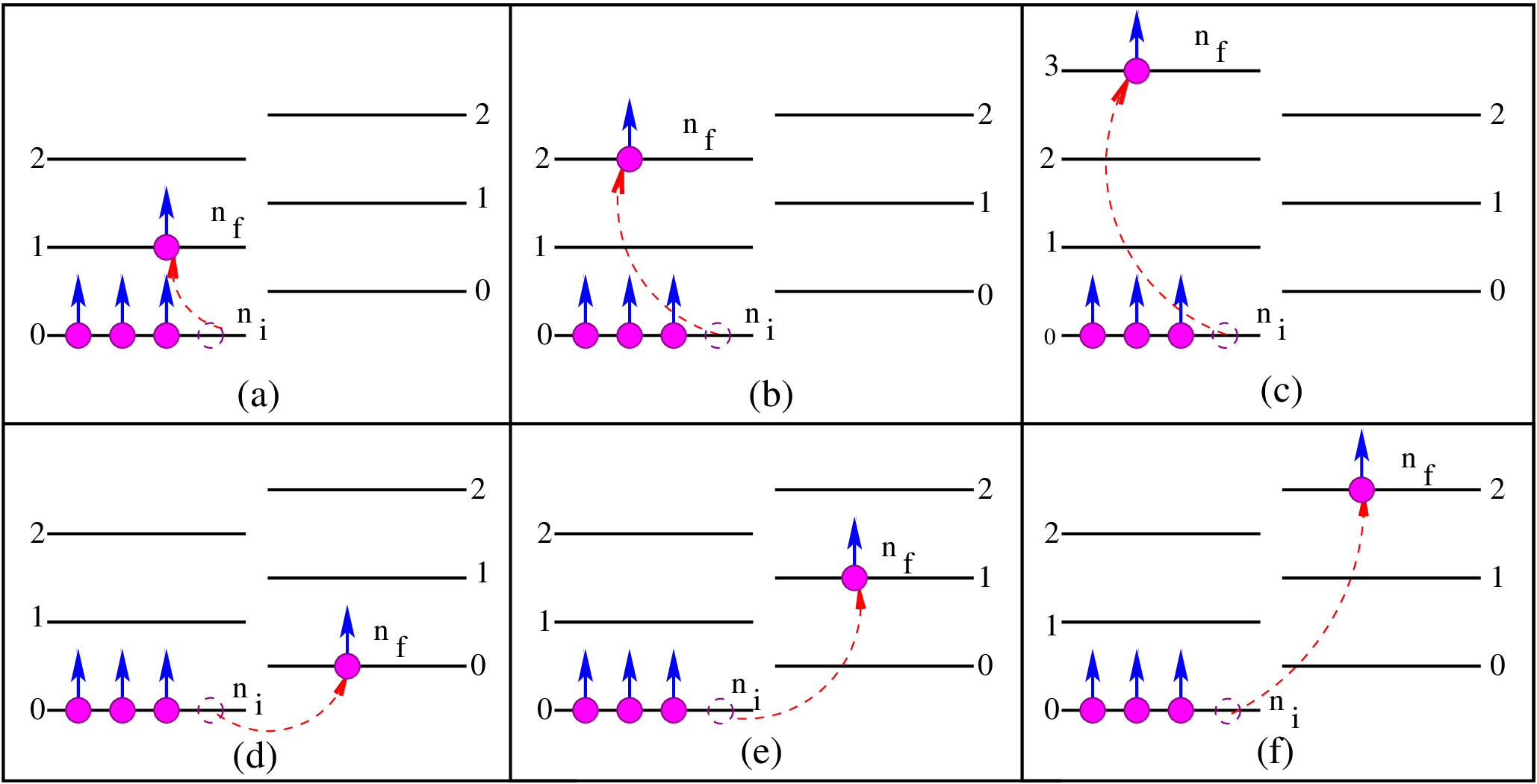}
\caption{
Schematic diagram of spin-conserving (a-c) and spin-reversed (d-f) excitons. In the left  panel, black straight lines represent spin-up, $\uparrow$, $\Lambda$-level, while, in the right panel they represent the down-spin, $\downarrow$, $\Lambda$-level. Possible excitations for $\nu=1/2$ are depicted, which follows for the other two filling fraction as well. Each block represents CF excitons (particle-hole pair) where dotted circle represents missing of particles from $n_i$-th $\Lambda$ level and that particle goes to $ n_f$-th $\Lambda$ level. Top panel blocks represent the possible spin-conserving excitations without change of spin state, while bottom row represents the spin-reversed excitation states. The lowest order exciton in the spin-conserving excitations are: (a) $|0>_\uparrow$ $\rightarrow$, $|1>_\uparrow$, (b) $|0>_\uparrow$ $ \rightarrow $ $|2>_\uparrow$, and (c) $|0>_\uparrow$ $ \rightarrow $ $|3>_\uparrow$. The lowest order exciton in the spin-reversed excitations are: (d) $|0>_\uparrow$ $\rightarrow$ $|0>_\downarrow$, (e) $|0>_\uparrow$ $\rightarrow$ $|1>_\downarrow$, and (f) $|0>_\uparrow$ $\rightarrow$ $|2>_\downarrow$.}
\label{fig1}
\end{figure*} 
We consider $N$ number of weakly interacting Bose atoms with mass $m$ rotating with an angular velocity $\Omega \hat{z}$ and confined in a 2D-isotropic harmonic trap potential in the 2D plane~\cite{Wilkin:2000}. The system Hamiltonian can be represented as,
\begin{eqnarray}
H&=& \sum_i \frac{1}{2m} (\vec{p}_i-\vec{A})^2 + (\omega-\Omega)L_z + V,  \
\label{H_boson}
\end{eqnarray}
where, $\vec{r_i}$ and $\vec{p_i}$ denote the position and momentum of the $i$-th atom respectively. $L_z$ is the $z$-component of angular momentum and $V$ is the interaction potential between the atoms and $\hbar\omega$ is the harmonic trap energy. The  Hamiltonian is expressed in terms of magnetic vector potential $\vec{A} = m \omega (-y, x)$ so that the corresponding magnetic field $\vec B = 2m\omega \hat z$ can be obtained along the z-axis, which indicates the equivalence to the Hamiltonian of a particle experiencing in presence of an effective magnetic field~\cite{Cooper:2001}. At high rotational frequency, as the system becomes sufficiently diluted, we consider that atoms get mainly confined to the lowest Landau levels. Note that the kinetic energy term is constant and thus has been ignored. To simplify it further we assume $\omega = \Omega$ which makes the Hamiltonian similar to that of the FQHE in the 2D fermionic system. The transformed Hamiltonian contains only the interacting bosons.\\ 

We have considered the atoms in the Bose gas are dipolar in nature with the magnetic dipole moment directed radially outward to the spherical geometry as a special case~\cite{Chung:2008}. Thus the interaction potential have the form as  
\begin{equation}
V = C_d \sum_{i<j=1}^N V(r_{ij})       
\end{equation}
where,
\begin{equation}
V(r_{ij})=\frac{\vec{p}_i \cdot \vec{p}_j - 3 (\vec{n}_{ij} \cdot \vec{p}_i) (\hat{n}_{ij} \cdot \vec{p}_j)}{r_{ij}^3}
\end{equation}
where, $C_d$ contains square of dipole moment of atoms and all other constants, $\hat{p}_i$ is the direction of the dipole moment of the $i^{\mbox{th}}$ atom, $\hat{n}_{ij}$ is the unit vector along the direction of connecting between the two atoms. On the spherical surface the dipole-dipole interaction between the pair of dipolar particles can be expressed as
\begin{equation}
  V(r_{ij}) = \frac{1}{r_{ij}^3} \left[ \vec{p}_i\cdot\vec{p}_j - 3 (\vec{p}_i\cdot \hat{r}_{ij})(\vec{p}_j\cdot \hat{r}_{ij})\right]
\end{equation}
We assume that all the dipoles get aligned along the perpendicular radial direction of the sphere, which implies, $\vec{p}_i = p \hat{r}_i$.  Using  $\vec{r}_{ij} = \vec{r}_i-\vec{r}_j$, the projection of the separation between the atoms on the position vector of the individual atom can be expressed as 
\begin{eqnarray}
  \vec{r}_{ij}\cdot \hat{r}_i = R (1-\hat{r}_i\cdot\hat{r}_j) \mbox{~~~~~~~~~(a)} \nonumber \\
  \vec{r}_{ij}\cdot \hat{r}_j = R (\hat{r}_i\cdot\hat{r}_j-1) \mbox{~~~~~~~~~(b)} \nonumber 
\end{eqnarray}
Further we express the  potential in the unit of $C_d$, which contains all the constant factors. With this potential takes the form as
\begin{eqnarray}
   V(r_{ij}) &=& \frac{1}{r_{ij}^3} \left[\frac{3R^2\; (1- \hat{r}_i\cdot\hat{r}_j)^2}{r_{ij}^2} +  \hat{r}_i\cdot\hat{r}_j    \right] \nonumber \\
             &=& \frac{1}{r_{ij}^3} \left[\frac{3\; (1- \hat{r}_i\cdot\hat{r}_j)^2}{(r_{ij}/R)^2} +  \hat{r}_i\cdot\hat{r}_j    \right] 
\end{eqnarray}
where, we use
\begin{eqnarray}
  \hat{r}_i\cdot\hat{r}_j = \sin \theta_i \; \sin\theta_j \; \cos(\phi_i - \phi_j)\; + \; \cos\theta_i \; \cos\theta_j \; .\nonumber
\end{eqnarray}
Now the two-body potential can be expressed as,
\begin{eqnarray}
  V(\vec{r}_i - \vec{r}_j) = \sum_m V_m P_m^{i,j} \nonumber
\end{eqnarray}
Here $P_m^{i,j}$ projects the two particles onto a state of relative angular momentum $m$. The parameters $V_m$, called Haldane pseudopotentials \cite{Simon:2007} are the energies of two particles in a state of relative angular momentum $m$.
In the next section, we present the many-body wave function for the rotating dipolar Bose gas that we have computed in the paper to determine the excitation spectra.
 \begin{figure}
 \centering
 \includegraphics[width=0.49\textwidth]{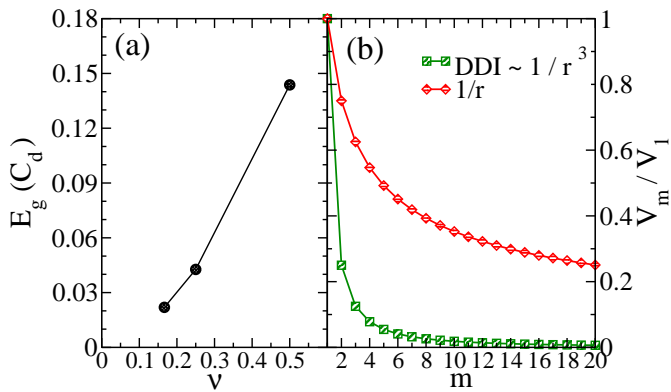}
 \caption{Ground state energy ($E_g$) per particle is plotted for three filling fractions at the left block (a). The pseudo-potential parameters ($V_m$) is calculated for dipole-dipole interaction as a function of relative angular momentum ($m$) of two particles. The ratio $V_m / V_1$ is plotted at the right block (b). The red curve which is for Coulomb-type potential~$1/r$ is included for comparison whereas the green one represents for DDI~$1/r^3$.}
 \label{Eg_Vm}
 \end{figure}
\section{Many-body wave-function and simulation details} \label{sec:waveFun}
Fractional quantum Hall effect (FQHE) is in general a topologically protected system with broken time-reversal symmetry~\cite{Gong:2014}. Although this phenomenon is independent of geometry, spherical geometry is considered to be ideal to investigate the bulk characteristics of this system due to its edgeless nature~\cite{Majumder:2009}. Following this we consider the composite fermion wave function in the typical spherical geometry \cite{Haldane:1983, Jain:2007, Haldane:1985, Dirac:1931} for all our numerical computations. We consider that that the $N$ number of correlated atoms exist on the 2D surface of a sphere with radius $R$. A magnetic monopole with strength $Q$ is located at the centre of the sphere which generates radial magnetic field with total magnetic flux given as $2Q\phi_0$ throughout the sphere's surface. Here, $\phi_0$ is the flux quanta. The radius of the sphere can be expressed as $R =\sqrt{Q} l_B$ (with $l_B=\sqrt{\frac{\hbar c}{eB}}$ is the magnetic length). This leads the effective flux of CFs as $2q = 2Q - p(N - 1)$. We choose the value of $Q$ in such a way that the state at $q$ is an integral of quantum Hall state with an integer filling as  $n$, which results the filling fraction $\nu = \frac{n}{n p + 1}$. With this the angular momentum ($L$) for a particle in the $n$-th Landau Level will be $n+Q-1$ \cite{Wu:1976, Wu:1977}. The single particle basis state is expressed in terms of the monopole harmonics given as $Y_{Q,n,m}$~\cite{Jain_Kamilla:1997} with $m$ as the $z$-component of angular momentum.

Next we present the ground state and excited state wave-functions for the FQH states at filling fraction $\nu$.
\\

{\bf A. Ground state wave-function:}
The ground state wave-function at filling fraction $\nu$ for $N$ Bose atoms is given by \cite{Wu:2013},
\begin{equation}
  \Psi_0 = J^{-1} P_{LLL} J^2 \; \Phi(\Omega_1, \Omega_2, \cdots \Omega_N) \;.
\end{equation}
Here, $\Omega_i \equiv \{\theta_i, \phi_i \}$ represents the angular position of $i$-th particle at the spherical surface and $\Phi$ denotes the Slater determinant of the fully filled $n$-number of $\Lambda$-levels of Bose-CFs. $P_{LLL}$ projects all the particles on  the LLL~\cite{Jain:1997} state, whereas, $J$ denotes the Jastrow factor which takes the form as
\begin{eqnarray}\label{eq:Jas}
  J = \prod _{i<j}^N (u_i v_j - u_j v_i)^p  
\end{eqnarray}
In the above equation (Eq.~\ref{eq:Jas}) we have used the spherical Spinor variables ($u,v$) which are given by
\begin{equation}
  u = \cos(\theta/2) e^{-i\phi/2} \mbox{~~~ and ~~~~} v = \sin (\theta/2) e^{i\phi/2} 
\end{equation}
where, $\theta$ and $\phi$ vary between $0$ to $\pi$ and $0$ to $2\pi$, respectively. The composite Fermion transformation maintains the overall wave-function symmetry of the system as $\Phi$ and $J$ both are antisymmetric in nature. \\

{\bf B. Wave-function for spin-conserving excitation:}
In the Spin-conserving excitation the transition from the filled $0$-th $\Lambda$ level to an empty $n_f\;$-th $\Lambda$ level takes place under composite transformation protocol in which only one type of the  spin component is involved~\cite{Majumder:2009, Majumder:2014}. For this configuration of the spin state the excited state wave-function can be expressed as
{\small{
\begin{equation}
  \Psi (L) = J^{-1} P_{LLL} J^2 \sum_{m_h} |m_h>\; <q, m_h; n_f+q,m_p|L,M>
\end{equation}}}
 where, $|m_h>$ is the Slater determinant of $N-1$ number of particles with a quasi-hole in the lowest $\Lambda$-level and one quasi-particle in the $n_f$-th $\Lambda$ level. Here, $<q, m_h;n_f+q,m_p|L,M>$ are the Clebsch-Gordan coefficients, $L$ is the total angular momentum, and $M$ is the $z$-component of the angular momentum. Further $m_p$ and $m_h$ are the angular momenta state of the quasi-particle and quasi-hole respectively. Spin-conserving excitations can be realized as one composite Fermion jumps from 0$\uparrow$ to the higher $\Lambda$-levels of the same spin. In this paper we have taken account of mainly three exciton states, which are represented as  $|0>_\uparrow$ $\rightarrow$ $|1>_\uparrow$;  $|0>_\uparrow$ $ \rightarrow  $ $|2>_\uparrow$ and $|0>_\uparrow$ $ \rightarrow  $ $|3>_\uparrow$. \\

{\bf C. Wave-function for spin-reversed excitation:}
In the spin-reversed excitation of $N$-particle system under composite Fermion transformation a transition takes from the filled $0$-th $\Lambda$ level to the other spin $n_f$-th $\Lambda$ level. The excited state wave function is obtained by taking the slater determinant of the $N-1$ number of particles ($|m_h>_{\small{N-1}}$) in one spin ($\uparrow$) state together with one single-particle state in another spin ($\downarrow$) state \cite{Indra:2020, Indra:2024}. We can express it as
\begin{eqnarray}
  \Psi (L) &=& J^{-1} P_{LLL} J^2 \sum_{m_h} |m_h>_{\small{N-1}}\; Y_{Q,n_f,m_p} \\ \nonumber
  &&<q, m_h; n_f+q,m_p|L,M> \;.
\end{eqnarray}
Here, $m_p$ and $m_h$ are the $z$-component momenta of the exciton state. The transition $|0>_\uparrow$ $\rightarrow$ $|0>_\downarrow$ is the conventional spin-wave excitation, whereas, $|0>_\uparrow$ $\rightarrow$ $|1>_\downarrow$ and $|0>_\uparrow$ $\rightarrow$ $|2>_\downarrow$ are the spin-flip (SF) excitations. These three lowest spin-reversed exciton states are considered in our study.

In Fig.~\ref{fig1} we show a schematic diagram of the formation of quasi-particles. Solid pink dots with blue arrow lines represent the composite fermions, whereas, solid dot represents the Bose atom with magnetic flux attachment which is shown with the arrow lines. Bose atoms can have odd number of flux quanta from the external magnetic field and thus can form composite Fermion particles which obey fermionic statistics. In our work, we have considered mainly the three flavors of CFs ($^1CF$, $^3CF$, $^5CF$) which have respectively one, three and five fluxes associated with them. In this configuration as one particle is being excited to higher LLs, the whole process gives rise to generation of an exciton. We have also schematically shown the excitation of the composite fermion corresponding to the spin conserved  and spin reversed configuration as shown in the panels (a-c) and (d-f) of Fig.~\ref{fig1} respectively.\\ 
\textbf {Simulation details}:
For all our simulation we have used the Metropolis Monte Carlo approach to evaluate the scalar inner products of various wave-functions and the Hamiltonian matrix elements. We typically perform $20$ Monte Carlo runs for a fixed $(N, Q)$. Each Monte Carlo run is done by averaging over $4\times10^6$ iterations for each angular momentum $L$. Since the blocks of $L$ are not coupled by the interaction, we can diagonalize them separately. To ensure accuracy, we have considered three different particle numbers $N = 75, 95, 125$ in our calculations and took the best fit to represent the results. We checked the energy spectra does not depend on the number of particles that denotes the thermodynamical behavior of the system.

\section{Collective excitation spectra}\label{sec:ExciSpec}
In this section, we present the numerical results of the collective excitation for the spin-conserving and spin-reversed excitation state as discussed in the previous section. Collective excitations provide the useful information about the different quantum phases and their stability with respect to the perturbation present in the system. While there are large number of works present for the collective excitation of the weakly interacting Bose gas, the similar analysis for the many-body interacting boson are few in number. There are some recent work that show the tremendous usefulness of this analysis to explain the characteristics and stability of different fractional quantum Hall states both in fermionic~\cite{Majumder:2014, Das:2017} and bosonic system \cite{Indra:2020, Indra:2024}. For rotating Bose gas, short-ranged contact interactions such as delta-function~\cite{Viefers:2008}, Pöschl-Teller type potential~\cite{Das:2018, Indra:2023} have been considered to model the excitation presents in the FQH states. However, for dipolar Bose gas such analysis is lacking. 

We begin our analysis by first showing the variation of the ground state energy per particle (in units of $C_d$) for three different filling fractions in Fig.~\ref{Eg_Vm} (a). We find that the energy increases with increase in the filling fractions. Just to make an comparison of our observation with other long range interaction, in Fig. \ref{Eg_Vm} (b) we show the ratio of pseudopotentials as a function of the relative momenta of two particles for both the DDI potential as well as Coulomb potential ($1/r$). Here we find that for DDI potential pseudopotential vanishes with higher relative momenta $m$, and becomes zero beyond around $m=10$. However, for Coulomb potential the energy decreases with increase of $m$ but attains some finite value even for higher $m$ ($m\sim 20$). At later section of the paper, we will establish a connection of the softening of the roton minima present in the excitation spectra of DDI potential with this particular feature.

In this paper we have considered the dipolar Bose atoms with strong dipole moment which interact through the usual dipole-dipole interaction. To avoid the numerical complexity, we choose the $z$-component of the total angular momentum ($M=0$) to be zero without any loss of generality. We have considered the possible collective spin-conserving and spin-reversed excitations for $\nu = 1/2 $ filling fraction [see Fig.~\ref{fig1}]. For the other two filling fractions also we consider similar types of excitations. In general we can have three possible lowest exciton for the rotating dipolar Bose gas considered here for both spin-conserving and spin-reversed excitation. However, as these states are not orthogonal to each other we perform the Gram-Schmidt orthonormalization procedure in the similar line as presented in~\cite{Mandal:2002} to orthogonalize low energy exciton states with a fixed angular momentum. \\  
 
 Let $\{\chi^i(L)\}$ be the orthonormalized states out of $\{\Psi (L)\}$.
The $i$-th excited mode of spectra with respect to ground-state energy is obtained by 
\begin{equation}
  \Delta^i(L) = \frac{<\chi^i_L| H |\chi^i_L>}{<\chi^i_L |\chi^i_L>} - \frac{<\Psi_0| H |\Psi_0 >}{<\Psi_0 | \Psi_0 > }
  \label{Del_L}
\end{equation}
The excitation energy is represented by the first term at the right hand side of the above equation (\ref{Del_L}) and the ground state energy by the latter one. The Hamiltonian of the system in equation (\ref{H_boson}) will be the interaction potential ($V$) only as described earlier. The bra-ket notation denotes the multidimensional integrations here, which are numerically computed by the Monte Carlo method.
\begin{figure}[!htp]
\centering
\includegraphics[width=0.48\textwidth]{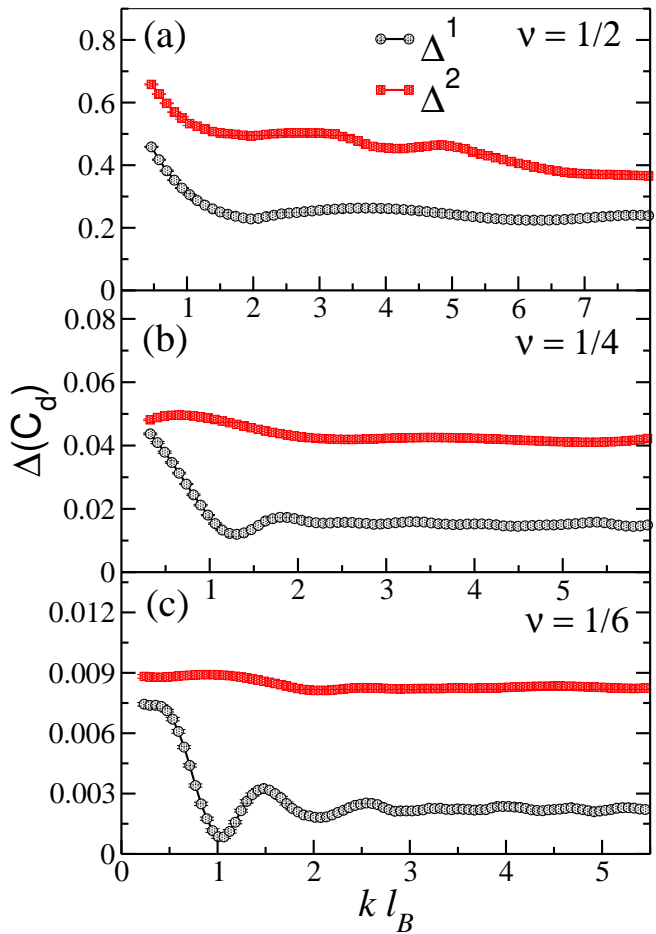}
\caption{Spin-conserving excitation spectra for different filling fraction (a) $\nu = 1/2$, (b) $\nu=1/4$, (c) $\nu=1/6$ as a function of wave number. Black colored curve denotes the fundamental lowest spectra and red colored one represents the next higher energy mode in each panel. The energies are in the units of $C_d$ and the wave number ($k$) is normalized with $l_B$. Here $k = L/R$, with $R$ is the radius of the sphere in the unit of magnetic length given by $l_B = \sqrt{\hbar c/eB}$.}
\label{CDE}
\end{figure}
\begin{figure}
\centering
\includegraphics[width=0.48\textwidth]{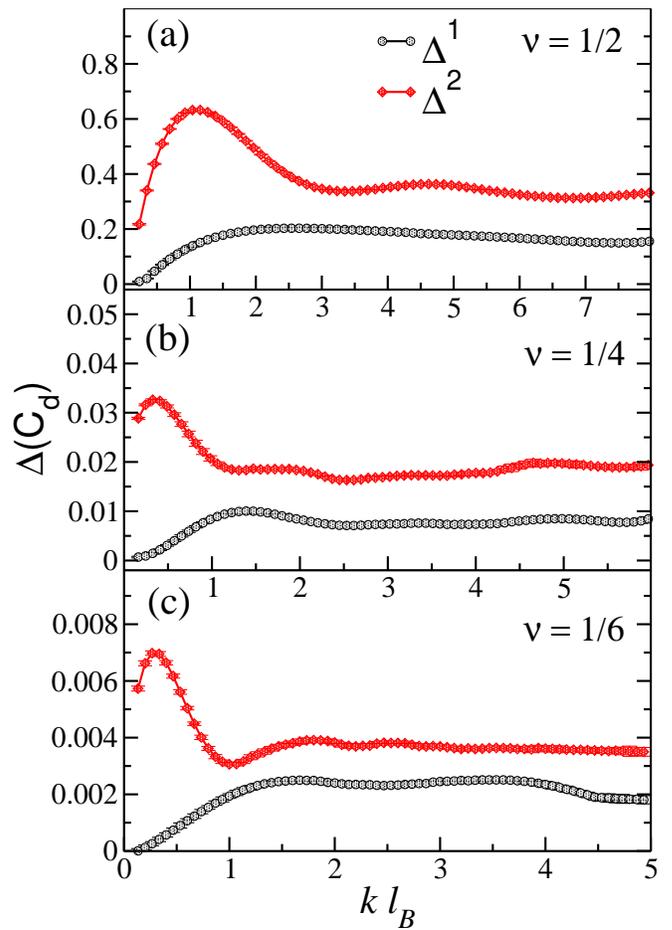}
\caption{Spin-reversed excitation spectra for the fully polarized state three different series of filling fraction (a) $\nu = 1/2$, (b)$\nu = 1/4$, and (c) $\nu= 1/6$. Black colored curve denotes the fundamental lowest spectra  and red colored one represents the next higher energy mode in each panel. The energies are given in the units of $C_d$, while  wave number  ($k$) is normalized with $l_B$.}
\label{SFE}
\end{figure} 
We have considered three different filling fractions $\nu = 1/2, 1/4, 1/6$ to compute the energy spectra ($\Delta$) for both the spin-conserving and spin-reversed excitations. In Fig.~\ref{CDE} we first discuss the results of the energy spectra for the lowest-order fundamental mode ($\Delta^1$) and the next higher-order excitation mode ($\Delta^2$) for the spin-conserving excitations. For this protocol we find the presence of a negative dispersion at low momenta value with the appearance of prominent roton-minima in the fundamental mode. The roton-minima appears to be  more prominent for the filling fraction $\nu=1/4, 1/6$ than that for $\nu=1/2$, which is consistent with the previous study with the Coulomb interaction~\cite{Indra:2023}. In general for the case of filling fraction $\nu=1/2$ as depicted in Fig.~\ref{CDE} (a) we find that at low momenta the fundamental mode ($\Delta^1$) exhibits  decreasing trend with momenta and beyond $k=2.0$ it becomes almost constant. However the next higher excited mode ($\Delta ^2$) also exhibits decreasing trend and similarly attains the constant value for larger momenta value. Interestingly at higher filling always there is existence of a finite energy gap between the two modes. For filling fraction $\nu=1/4$  fundamental mode exhibits decreasing trend upon increase of momenta and also exhibits the roton-minima at $k=1.3$ [See Fig.~\ref{CDE} (b)]. For further lowering the filling fraction to $\nu=1/6$, we find that the two roton-minima in the fundamental mode appears respectively at $k=1.0$ and $k=2.0$, whereas, the higher excitation mode remains constant with the momenta [See Fig.~\ref{CDE} (c)]. Overall we notice that the gap between the fundamental mode and higher excited mode for zero momenta ($k=0$) decreases upon  decrease of the filling fractions with appearance of  gap closing for lower filling fraction. Also the roton-minima gets softened and further show a shifting towards lower momenta value upon decrease of filling fraction.

In case of spin-reversed excitation as depicted in Fig.~\ref{SFE}, we find that the fundamental modes show positive dispersion (spin-wave nature) at low momenta, which agrees with the results of the short-range interaction as presented in~\cite{Indra:2018}. However, the higher excited mode  exhibits different nature that observed for the spin-conserving protocols. For filling fraction $\nu=1/2$ as illustrated in Fig.~\ref{SFE} (a), we find that the fundamental mode shows increasing trend with increase of momenta and beyond $k=2.5$ they attain the constant. However, higher excited mode shows increasing behavior initially with the increase of momenta which further attains constant at higher momenta. Similarly for $\nu=1/4$, the fundamental mode becomes constant beyond $k=2.0$. However, for $\nu=1/6$, as shown in Fig.~\ref{SFE} (c) the fundamental mode attains constant beyond $k=1.5$ while the higher excited mode follows same nature as those observed for $\nu=1/4$. The gap between two modes decreases upon lowering the  filling fraction. 

After discussing the excitation spectra in the following section now we move our focus on identifying the excitation mode which will be more visible in the experiment. 
\section{Spectral weight}\label{sec:SpecWeight}
One important issue related to the excitation analysis presented in this work concerns its experimental realization. In general, the feasibility of observing such excitations can be assessed by analyzing the spectral weight associated with each momentum component of the excitation spectrum at different filling fractions. The spectral weight (SW) quantifies how strongly a particular excitation mode contributes to the system’s overall response. It serves as a key indicator of the strength or visibility of a mode when probed experimentally—particularly through techniques such as Raman scattering. The spectral weight ($S_L$) is defined as~\cite{Majumder:2009}
\begin{equation}\label{eq:SW}
  S_L = \frac{1}{N} |<\chi_L| \rho_L |\Psi_0>|^2
\end{equation}
where $|\Psi_0>$ is the ground state wave function, $\rho_L$ is the number density given by
\begin{eqnarray}
  \rho_L = \sum_{k=1}^N Y_{L,0} (\theta_k, \phi_k)
\end{eqnarray}
with $Y_{L,0} (\Omega_k)$ is the single particle monopole harmonics.
$|\chi_L>$ is the Gram-Schmidt Orthonormalized wave function given by 
\begin{eqnarray}
  |\chi_L> = \sum_l C_L^{l} |\Psi_L^l> \;\;;
\end{eqnarray}
where, $|\Psi_L^l>$ is the normalized exciton wave function, $l$ denotes the transition between $n_i \rightarrow n_f$.

\begin{figure}
 \centering
 \includegraphics[width=0.5\textwidth]{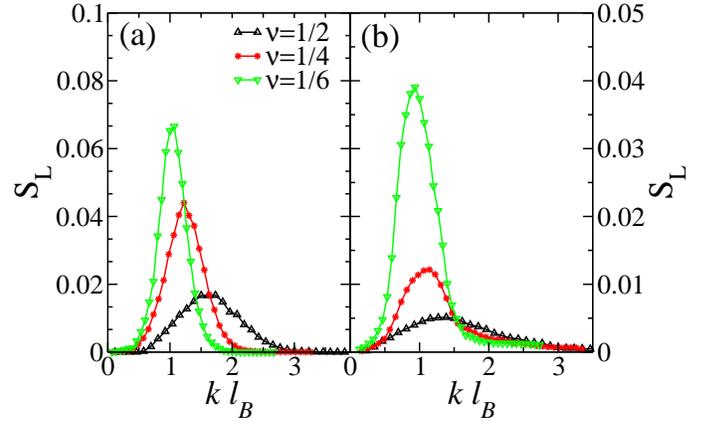}
 \caption{Spectral weight ($S_L$) for all the fundamental mode of excitations as a function of linear momentum $k = L/R$ for (a) spin-conserving and (b) spin-reversed excitation for different filling fraction $\nu =1/2, 1/4, 1/6$. The peak of spectral weight shifts towards the lower momenta upon decrease of the filling fraction.}
 \label{SW}
 \end{figure}

In order to  obtain the spectral weight, one needs to determine the normalized wave function $|\Psi_L^l>$ and the coefficients $C_L^{l}$. In general the lowest-energy mode has the highest spectral weight, and other higher-energy modes are negligible. We consider the SW only for the fundamental mode of excitation. In Fig.~\ref{SW}, we depict the spectral weight for the fundamental modes of the excitation spectra for different filling fractions. We find that for the spin-conserving excitations the spectral weight tends to exhibit higher value for those in the spin-reversed excitations. Interestingly we find that among the two cases, the filling fraction $\nu=1/6$ has the highest SW, while $\nu=1/2$ has the lowest SW, and $\nu=1/4$ has an intermediate value. We find that while the peak of the spectral weight shifts towards the lower momenta upon decreasing the filling fraction for the spin-conserving excitation (panel (a)), the momenta at which the SW peaks does not show much change for the spin-reversed excitation (panel (b)).  
\section{Conclusions}\label{sec:conclusion}
In this work, we have investigated the neutral collective excitation of rapidly rotating ultra-cold Bose atoms, which serve as an atomic analog of electrons in different FQHE \cite{Tsui_PRB:1982, Stormer:1983}, subjected to a strong magnetic field, and cooled at a low temperature. We have considered pseudo-spin $1/2$ Bose atoms having small scattering lengths where DDI becomes more important than the contact interaction. We assume that each of the Bose atoms with a dipole moment oriented perpendicular to the trapping potential and they interact with each other via long-ranged dipole-dipole interaction. We have taken into account the lowest possible excitations in our study and calculated the excitation spectra for the three fractions of the first Jain series $\nu = 1/2, 1/4, 1/6$. The lowest-order fundamental mode and the next higher-energy mode of excitation spectra are presented for each of the three fractions. We observed the nature of the excitation spectra follows the same pattern as observed in case of short-ranged contact interaction~\cite{Indra:2024}. Additionally here we added the spectra for next higher exciton mode and observed that in both the spin conserving and spin-reversed excitation the gap between the low lying spectrum decreases significantly with the decrease in the filling fraction. Further we compute the spectral weight for the fundamental mode of excitation, that shows a significant increment in the spectral intensity with decrease with the filling fraction. We can conclude that in an inelastic resonant Raman scattering experiment \cite{Pinczuk:1996}, collective excitation of $\nu=1/6$ will be more prominent than $\nu=1/4, 1/2$. 

The results presented in the work provide valuable insights into the nature of collective excitations in this strongly correlated regime, highlighting the influence of dipolar interactions and rotation-induced effects on the excitation spectrum. These findings may contribute to a deeper understanding of the interplay between rotation, quantum correlations, and many-body physics in dipolar Bose systems for the experimental achievement.

\section{Acknowledgements}
MI acknowledges the Institute Post-Doctoral Fellowship (IPDF) fund IIT Guwahati for her financial support.
\bibliography{reference.bib} 

\begin{thebibliography}{77}%
\makeatletter
\providecommand \@ifxundefined [1]{%
 \@ifx{#1\undefined}
}%
\providecommand \@ifnum [1]{%
 \ifnum #1\expandafter \@firstoftwo
 \else \expandafter \@secondoftwo
 \fi
}%
\providecommand \@ifx [1]{%
 \ifx #1\expandafter \@firstoftwo
 \else \expandafter \@secondoftwo
 \fi
}%
\providecommand \natexlab [1]{#1}%
\providecommand \enquote  [1]{``#1''}%
\providecommand \bibnamefont  [1]{#1}%
\providecommand \bibfnamefont [1]{#1}%
\providecommand \citenamefont [1]{#1}%
\providecommand \href@noop [0]{\@secondoftwo}%
\providecommand \href [0]{\begingroup \@sanitize@url \@href}%
\providecommand \@href[1]{\@@startlink{#1}\@@href}%
\providecommand \@@href[1]{\endgroup#1\@@endlink}%
\providecommand \@sanitize@url [0]{\catcode `\\12\catcode `\$12\catcode `\&12\catcode `\#12\catcode `\^12\catcode `\_12\catcode `\%12\relax}%
\providecommand \@@startlink[1]{}%
\providecommand \@@endlink[0]{}%
\providecommand \url  [0]{\begingroup\@sanitize@url \@url }%
\providecommand \@url [1]{\endgroup\@href {#1}{\urlprefix }}%
\providecommand \urlprefix  [0]{URL }%
\providecommand \Eprint [0]{\href }%
\providecommand \doibase [0]{https://doi.org/}%
\providecommand \selectlanguage [0]{\@gobble}%
\providecommand \bibinfo  [0]{\@secondoftwo}%
\providecommand \bibfield  [0]{\@secondoftwo}%
\providecommand \translation [1]{[#1]}%
\providecommand \BibitemOpen [0]{}%
\providecommand \bibitemStop [0]{}%
\providecommand \bibitemNoStop [0]{.\EOS\space}%
\providecommand \EOS [0]{\spacefactor3000\relax}%
\providecommand \BibitemShut  [1]{\csname bibitem#1\endcsname}%
\let\auto@bib@innerbib\@empty
\bibitem [{\citenamefont {Davis}\ \emph {et~al.}(1995)\citenamefont {Davis}, \citenamefont {Mewes}, \citenamefont {Andrews}, \citenamefont {van Druten}, \citenamefont {Durfee}, \citenamefont {Kurn},\ and\ \citenamefont {Ketterle}}]{Davis:1995}%
  \BibitemOpen
  \bibfield  {author} {\bibinfo {author} {\bibfnamefont {K.~B.}\ \bibnamefont {Davis}}, \bibinfo {author} {\bibfnamefont {M.-O.}\ \bibnamefont {Mewes}}, \bibinfo {author} {\bibfnamefont {M.~R.}\ \bibnamefont {Andrews}}, \bibinfo {author} {\bibfnamefont {N.~J.}\ \bibnamefont {van Druten}}, \bibinfo {author} {\bibfnamefont {D.~S.}\ \bibnamefont {Durfee}}, \bibinfo {author} {\bibfnamefont {D.}~\bibnamefont {Kurn}},\ and\ \bibinfo {author} {\bibfnamefont {W.}~\bibnamefont {Ketterle}},\ }\bibfield  {title} {\bibinfo {title} {Bose-einstein condensation in a gas of sodium atoms},\ }\href {https://doi.org/10.1103/PhysRevLett.75.3969} {\bibfield  {journal} {\bibinfo  {journal} {Physical review letters}\ }\textbf {\bibinfo {volume} {75}},\ \bibinfo {pages} {3969} (\bibinfo {year} {1995})}\BibitemShut {NoStop}%
\bibitem [{\citenamefont {Anderson}\ \emph {et~al.}(1995)\citenamefont {Anderson}, \citenamefont {Ensher}, \citenamefont {Matthews}, \citenamefont {Wieman},\ and\ \citenamefont {Cornell}}]{Anderson:1995}%
  \BibitemOpen
  \bibfield  {author} {\bibinfo {author} {\bibfnamefont {M.~H.}\ \bibnamefont {Anderson}}, \bibinfo {author} {\bibfnamefont {J.~R.}\ \bibnamefont {Ensher}}, \bibinfo {author} {\bibfnamefont {M.~R.}\ \bibnamefont {Matthews}}, \bibinfo {author} {\bibfnamefont {C.~E.}\ \bibnamefont {Wieman}},\ and\ \bibinfo {author} {\bibfnamefont {E.~A.}\ \bibnamefont {Cornell}},\ }\bibfield  {title} {\bibinfo {title} {Observation of bose-einstein condensation in a dilute atomic vapor},\ }\href {https://doi.org/10.1126/science.269.5221.198} {\bibfield  {journal} {\bibinfo  {journal} {science}\ }\textbf {\bibinfo {volume} {269}},\ \bibinfo {pages} {198} (\bibinfo {year} {1995})}\BibitemShut {NoStop}%
\bibitem [{\citenamefont {Tang}\ \emph {et~al.}(2018)\citenamefont {Tang}, \citenamefont {Kao}, \citenamefont {Li},\ and\ \citenamefont {Lev}}]{Tang:2018}%
  \BibitemOpen
  \bibfield  {author} {\bibinfo {author} {\bibfnamefont {Y.}~\bibnamefont {Tang}}, \bibinfo {author} {\bibfnamefont {W.}~\bibnamefont {Kao}}, \bibinfo {author} {\bibfnamefont {K.-Y.}\ \bibnamefont {Li}},\ and\ \bibinfo {author} {\bibfnamefont {B.~L.}\ \bibnamefont {Lev}},\ }\bibfield  {title} {\bibinfo {title} {Tuning the dipole-dipole interaction in a quantum gas with a rotating magnetic field},\ }\href {https://doi.org/10.1103/PhysRevLett.120.230401} {\bibfield  {journal} {\bibinfo  {journal} {Physical review letters}\ }\textbf {\bibinfo {volume} {120}},\ \bibinfo {pages} {230401} (\bibinfo {year} {2018})}\BibitemShut {NoStop}%
\bibitem [{\citenamefont {Regnault}\ and\ \citenamefont {Jolicoeur}(2003)}]{Regnault:2003}%
  \BibitemOpen
  \bibfield  {author} {\bibinfo {author} {\bibfnamefont {N.}~\bibnamefont {Regnault}}\ and\ \bibinfo {author} {\bibfnamefont {T.}~\bibnamefont {Jolicoeur}},\ }\bibfield  {title} {\bibinfo {title} {Quantum hall fractions in rotating bose-einstein condensates},\ }\href {https://doi.org/10.1103/PhysRevLett.91.030402} {\bibfield  {journal} {\bibinfo  {journal} {Physical review letters}\ }\textbf {\bibinfo {volume} {91}},\ \bibinfo {pages} {030402} (\bibinfo {year} {2003})}\BibitemShut {NoStop}%
\bibitem [{\citenamefont {Lu}\ \emph {et~al.}(2011)\citenamefont {Lu}, \citenamefont {Burdick}, \citenamefont {Youn},\ and\ \citenamefont {Lev}}]{Lu:2011}%
  \BibitemOpen
  \bibfield  {author} {\bibinfo {author} {\bibfnamefont {M.}~\bibnamefont {Lu}}, \bibinfo {author} {\bibfnamefont {N.~Q.}\ \bibnamefont {Burdick}}, \bibinfo {author} {\bibfnamefont {S.~H.}\ \bibnamefont {Youn}},\ and\ \bibinfo {author} {\bibfnamefont {B.~L.}\ \bibnamefont {Lev}},\ }\bibfield  {title} {\bibinfo {title} {Strongly dipolar bose-einstein condensate of dysprosium},\ }\href {https://doi.org/PhysRevLett.107.190401} {\bibfield  {journal} {\bibinfo  {journal} {Physical review letters}\ }\textbf {\bibinfo {volume} {107}},\ \bibinfo {pages} {190401} (\bibinfo {year} {2011})}\BibitemShut {NoStop}%
\bibitem [{\citenamefont {Trautmann}\ \emph {et~al.}(2018)\citenamefont {Trautmann}, \citenamefont {Ilzh{\"o}fer}, \citenamefont {Durastante}, \citenamefont {Politi}, \citenamefont {Sohmen}, \citenamefont {Mark},\ and\ \citenamefont {Ferlaino}}]{Trautmann:2018}%
  \BibitemOpen
  \bibfield  {author} {\bibinfo {author} {\bibfnamefont {A.}~\bibnamefont {Trautmann}}, \bibinfo {author} {\bibfnamefont {P.}~\bibnamefont {Ilzh{\"o}fer}}, \bibinfo {author} {\bibfnamefont {G.}~\bibnamefont {Durastante}}, \bibinfo {author} {\bibfnamefont {C.}~\bibnamefont {Politi}}, \bibinfo {author} {\bibfnamefont {M.}~\bibnamefont {Sohmen}}, \bibinfo {author} {\bibfnamefont {M.}~\bibnamefont {Mark}},\ and\ \bibinfo {author} {\bibfnamefont {F.}~\bibnamefont {Ferlaino}},\ }\bibfield  {title} {\bibinfo {title} {Dipolar quantum mixtures of erbium and dysprosium atoms},\ }\href {https://doi.org/10.1103/PhysRevLett.121.213601} {\bibfield  {journal} {\bibinfo  {journal} {Physical Review Letters}\ }\textbf {\bibinfo {volume} {121}},\ \bibinfo {pages} {213601} (\bibinfo {year} {2018})}\BibitemShut {NoStop}%
\bibitem [{\citenamefont {Yamamoto}\ and\ \citenamefont {Takahashi}(2016)}]{Yamamoto:2016}%
  \BibitemOpen
  \bibfield  {author} {\bibinfo {author} {\bibfnamefont {Y.}~\bibnamefont {Yamamoto}}\ and\ \bibinfo {author} {\bibfnamefont {Y.}~\bibnamefont {Takahashi}},\ }\bibfield  {title} {\bibinfo {title} {Bose-einstein condensation: A platform for quantum simulation experiments},\ }in\ \href {https://doi.org/10.1007/978-4-431-55756-2#page=267} {\emph {\bibinfo {booktitle} {Principles and Methods of Quantum Information Technologies}}}\ (\bibinfo  {publisher} {Springer},\ \bibinfo {year} {2016})\ pp.\ \bibinfo {pages} {265--307}\BibitemShut {NoStop}%
\bibitem [{\citenamefont {Lieb}\ and\ \citenamefont {Liniger}(1963)}]{Lieb:1963}%
  \BibitemOpen
  \bibfield  {author} {\bibinfo {author} {\bibfnamefont {E.~H.}\ \bibnamefont {Lieb}}\ and\ \bibinfo {author} {\bibfnamefont {W.}~\bibnamefont {Liniger}},\ }\bibfield  {title} {\bibinfo {title} {Exact analysis of an interacting bose gas. i. the general solution and the ground state},\ }\href {https://doi.org/10.1103/PhysRev.130.1605} {\bibfield  {journal} {\bibinfo  {journal} {Physical Review}\ }\textbf {\bibinfo {volume} {130}},\ \bibinfo {pages} {1605} (\bibinfo {year} {1963})}\BibitemShut {NoStop}%
\bibitem [{\citenamefont {Rogel-Salazar}(2013)}]{Rogel:2013}%
  \BibitemOpen
  \bibfield  {author} {\bibinfo {author} {\bibfnamefont {J.}~\bibnamefont {Rogel-Salazar}},\ }\bibfield  {title} {\bibinfo {title} {The gross--pitaevskii equation and bose--einstein condensates},\ }\href {https://doi.org/10.1088/0143-0807/34/2/247} {\bibfield  {journal} {\bibinfo  {journal} {European Journal of Physics}\ }\textbf {\bibinfo {volume} {34}},\ \bibinfo {pages} {247} (\bibinfo {year} {2013})}\BibitemShut {NoStop}%
\bibitem [{\citenamefont {Griesmaier}\ \emph {et~al.}(2005)\citenamefont {Griesmaier}, \citenamefont {Werner}, \citenamefont {Hensler}, \citenamefont {Stuhler},\ and\ \citenamefont {Pfau}}]{Griesmaier:2005}%
  \BibitemOpen
  \bibfield  {author} {\bibinfo {author} {\bibfnamefont {A.}~\bibnamefont {Griesmaier}}, \bibinfo {author} {\bibfnamefont {J.}~\bibnamefont {Werner}}, \bibinfo {author} {\bibfnamefont {S.}~\bibnamefont {Hensler}}, \bibinfo {author} {\bibfnamefont {J.}~\bibnamefont {Stuhler}},\ and\ \bibinfo {author} {\bibfnamefont {T.}~\bibnamefont {Pfau}},\ }\bibfield  {title} {\bibinfo {title} {Bose-einstein condensation of chromium},\ }\href {https://doi.org/10.1103/PhysRevLett.94.160401} {\bibfield  {journal} {\bibinfo  {journal} {Physical Review Letters}\ }\textbf {\bibinfo {volume} {94}},\ \bibinfo {pages} {160401} (\bibinfo {year} {2005})}\BibitemShut {NoStop}%
\bibitem [{\citenamefont {Stuhler}\ \emph {et~al.}(2005)\citenamefont {Stuhler}, \citenamefont {Griesmaier}, \citenamefont {Koch}, \citenamefont {Fattori}, \citenamefont {Pfau}, \citenamefont {Giovanazzi}, \citenamefont {Pedri},\ and\ \citenamefont {Santos}}]{Stuhler:2005}%
  \BibitemOpen
  \bibfield  {author} {\bibinfo {author} {\bibfnamefont {J.}~\bibnamefont {Stuhler}}, \bibinfo {author} {\bibfnamefont {A.}~\bibnamefont {Griesmaier}}, \bibinfo {author} {\bibfnamefont {T.}~\bibnamefont {Koch}}, \bibinfo {author} {\bibfnamefont {M.}~\bibnamefont {Fattori}}, \bibinfo {author} {\bibfnamefont {T.}~\bibnamefont {Pfau}}, \bibinfo {author} {\bibfnamefont {S.}~\bibnamefont {Giovanazzi}}, \bibinfo {author} {\bibfnamefont {P.}~\bibnamefont {Pedri}},\ and\ \bibinfo {author} {\bibfnamefont {L.}~\bibnamefont {Santos}},\ }\bibfield  {title} {\bibinfo {title} {Observation of dipole-dipole interaction in a degenerate quantum gas},\ }\href {https://doi.org/10.1103/PhysRevLett.95.150406} {\bibfield  {journal} {\bibinfo  {journal} {Physical Review Letters}\ }\textbf {\bibinfo {volume} {95}},\ \bibinfo {pages} {150406} (\bibinfo {year} {2005})}\BibitemShut {NoStop}%
\bibitem [{\citenamefont {Lahaye}\ \emph {et~al.}(2007)\citenamefont {Lahaye}, \citenamefont {Koch}, \citenamefont {Fr{\"o}hlich}, \citenamefont {Fattori}, \citenamefont {Metz}, \citenamefont {Griesmaier}, \citenamefont {Giovanazzi},\ and\ \citenamefont {Pfau}}]{Lahaye:2007}%
  \BibitemOpen
  \bibfield  {author} {\bibinfo {author} {\bibfnamefont {T.}~\bibnamefont {Lahaye}}, \bibinfo {author} {\bibfnamefont {T.}~\bibnamefont {Koch}}, \bibinfo {author} {\bibfnamefont {B.}~\bibnamefont {Fr{\"o}hlich}}, \bibinfo {author} {\bibfnamefont {M.}~\bibnamefont {Fattori}}, \bibinfo {author} {\bibfnamefont {J.}~\bibnamefont {Metz}}, \bibinfo {author} {\bibfnamefont {A.}~\bibnamefont {Griesmaier}}, \bibinfo {author} {\bibfnamefont {S.}~\bibnamefont {Giovanazzi}},\ and\ \bibinfo {author} {\bibfnamefont {T.}~\bibnamefont {Pfau}},\ }\bibfield  {title} {\bibinfo {title} {Strong dipolar effects in a quantum ferrofluid},\ }\href {https://doi.org/10.1038/nature06036} {\bibfield  {journal} {\bibinfo  {journal} {Nature}\ }\textbf {\bibinfo {volume} {448}},\ \bibinfo {pages} {672} (\bibinfo {year} {2007})}\BibitemShut {NoStop}%
\bibitem [{\citenamefont {Olson}\ \emph {et~al.}(2013)\citenamefont {Olson}, \citenamefont {Whitenack},\ and\ \citenamefont {Chen}}]{Olson:2013}%
  \BibitemOpen
  \bibfield  {author} {\bibinfo {author} {\bibfnamefont {A.~J.}\ \bibnamefont {Olson}}, \bibinfo {author} {\bibfnamefont {D.~L.}\ \bibnamefont {Whitenack}},\ and\ \bibinfo {author} {\bibfnamefont {Y.~P.}\ \bibnamefont {Chen}},\ }\bibfield  {title} {\bibinfo {title} {Effects of magnetic dipole-dipole interactions in atomic bose-einstein condensates with tunable s-wave interactions},\ }\href {https://doi.org/10.1103/PhysRevA.88.043609} {\bibfield  {journal} {\bibinfo  {journal} {Physical Review A—Atomic, Molecular, and Optical Physics}\ }\textbf {\bibinfo {volume} {88}},\ \bibinfo {pages} {043609} (\bibinfo {year} {2013})}\BibitemShut {NoStop}%
\bibitem [{\citenamefont {Xi}\ and\ \citenamefont {Saito}(2016)}]{Xi:2016}%
  \BibitemOpen
  \bibfield  {author} {\bibinfo {author} {\bibfnamefont {K.-T.}\ \bibnamefont {Xi}}\ and\ \bibinfo {author} {\bibfnamefont {H.}~\bibnamefont {Saito}},\ }\bibfield  {title} {\bibinfo {title} {Droplet formation in a bose-einstein condensate with strong dipole-dipole interaction},\ }\href {https://doi.org/10.1103/PhysRevA.93.011604} {\bibfield  {journal} {\bibinfo  {journal} {Physical Review A}\ }\textbf {\bibinfo {volume} {93}},\ \bibinfo {pages} {011604} (\bibinfo {year} {2016})}\BibitemShut {NoStop}%
\bibitem [{\citenamefont {Bismut}\ \emph {et~al.}(2010)\citenamefont {Bismut}, \citenamefont {Pasquiou}, \citenamefont {Mar{\'e}chal}, \citenamefont {Pedri}, \citenamefont {Vernac}, \citenamefont {Gorceix},\ and\ \citenamefont {Laburthe-Tolra}}]{Bismut:2010}%
  \BibitemOpen
  \bibfield  {author} {\bibinfo {author} {\bibfnamefont {G.}~\bibnamefont {Bismut}}, \bibinfo {author} {\bibfnamefont {B.}~\bibnamefont {Pasquiou}}, \bibinfo {author} {\bibfnamefont {E.}~\bibnamefont {Mar{\'e}chal}}, \bibinfo {author} {\bibfnamefont {P.}~\bibnamefont {Pedri}}, \bibinfo {author} {\bibfnamefont {L.}~\bibnamefont {Vernac}}, \bibinfo {author} {\bibfnamefont {O.}~\bibnamefont {Gorceix}},\ and\ \bibinfo {author} {\bibfnamefont {B.}~\bibnamefont {Laburthe-Tolra}},\ }\bibfield  {title} {\bibinfo {title} {Collective excitations of a dipolar bose-einstein condensate},\ }\href {https://doi.org/10.1103/PhysRevLett.105.040404} {\bibfield  {journal} {\bibinfo  {journal} {Physical review letters}\ }\textbf {\bibinfo {volume} {105}},\ \bibinfo {pages} {040404} (\bibinfo {year} {2010})}\BibitemShut {NoStop}%
\bibitem [{\citenamefont {Lahaye}\ \emph {et~al.}(2009)\citenamefont {Lahaye}, \citenamefont {Menotti}, \citenamefont {Santos}, \citenamefont {Lewenstein},\ and\ \citenamefont {Pfau}}]{Lahaye:2009}%
  \BibitemOpen
  \bibfield  {author} {\bibinfo {author} {\bibfnamefont {T.}~\bibnamefont {Lahaye}}, \bibinfo {author} {\bibfnamefont {C.}~\bibnamefont {Menotti}}, \bibinfo {author} {\bibfnamefont {L.}~\bibnamefont {Santos}}, \bibinfo {author} {\bibfnamefont {M.}~\bibnamefont {Lewenstein}},\ and\ \bibinfo {author} {\bibfnamefont {T.}~\bibnamefont {Pfau}},\ }\bibfield  {title} {\bibinfo {title} {The physics of dipolar bosonic quantum gases},\ }\href {https://doi.org/10.1088/0034-4885/72/12/126401} {\bibfield  {journal} {\bibinfo  {journal} {Reports on Progress in Physics}\ }\textbf {\bibinfo {volume} {72}},\ \bibinfo {pages} {126401} (\bibinfo {year} {2009})}\BibitemShut {NoStop}%
\bibitem [{\citenamefont {Chomaz}\ \emph {et~al.}(2018)\citenamefont {Chomaz}, \citenamefont {van Bijnen}, \citenamefont {Petter}, \citenamefont {Faraoni}, \citenamefont {Baier}, \citenamefont {Becher}, \citenamefont {Mark}, \citenamefont {Waechtler}, \citenamefont {Santos},\ and\ \citenamefont {Ferlaino}}]{Chomaz:2018}%
  \BibitemOpen
  \bibfield  {author} {\bibinfo {author} {\bibfnamefont {L.}~\bibnamefont {Chomaz}}, \bibinfo {author} {\bibfnamefont {R.~M.}\ \bibnamefont {van Bijnen}}, \bibinfo {author} {\bibfnamefont {D.}~\bibnamefont {Petter}}, \bibinfo {author} {\bibfnamefont {G.}~\bibnamefont {Faraoni}}, \bibinfo {author} {\bibfnamefont {S.}~\bibnamefont {Baier}}, \bibinfo {author} {\bibfnamefont {J.~H.}\ \bibnamefont {Becher}}, \bibinfo {author} {\bibfnamefont {M.~J.}\ \bibnamefont {Mark}}, \bibinfo {author} {\bibfnamefont {F.}~\bibnamefont {Waechtler}}, \bibinfo {author} {\bibfnamefont {L.}~\bibnamefont {Santos}},\ and\ \bibinfo {author} {\bibfnamefont {F.}~\bibnamefont {Ferlaino}},\ }\bibfield  {title} {\bibinfo {title} {Observation of roton mode population in a dipolar quantum gas},\ }\href {https://doi.org/10.1038/s41567-018-0054-7} {\bibfield  {journal} {\bibinfo  {journal} {Nature physics}\ }\textbf {\bibinfo {volume} {14}},\ \bibinfo {pages} {442} (\bibinfo {year} {2018})}\BibitemShut {NoStop}%
\bibitem [{\citenamefont {Lyu}\ \emph {et~al.}(2022)\citenamefont {Lyu}, \citenamefont {Zhang},\ and\ \citenamefont {Busch}}]{Lyu:2022}%
  \BibitemOpen
  \bibfield  {author} {\bibinfo {author} {\bibfnamefont {H.}~\bibnamefont {Lyu}}, \bibinfo {author} {\bibfnamefont {Y.}~\bibnamefont {Zhang}},\ and\ \bibinfo {author} {\bibfnamefont {T.}~\bibnamefont {Busch}},\ }\bibfield  {title} {\bibinfo {title} {Detection of roton and phonon excitations in a spin-orbit-coupled bose-einstein condensate with a moving barrier},\ }\href {https://doi.org/10.1103/PhysRevA.106.013302} {\bibfield  {journal} {\bibinfo  {journal} {Physical Review A}\ }\textbf {\bibinfo {volume} {106}},\ \bibinfo {pages} {013302} (\bibinfo {year} {2022})}\BibitemShut {NoStop}%
\bibitem [{\citenamefont {Hung}\ \emph {et~al.}(2012)\citenamefont {Hung}, \citenamefont {Song}, \citenamefont {Chen}, \citenamefont {Ma}, \citenamefont {Xue},\ and\ \citenamefont {Wu}}]{Hung:2012}%
  \BibitemOpen
  \bibfield  {author} {\bibinfo {author} {\bibfnamefont {H.-H.}\ \bibnamefont {Hung}}, \bibinfo {author} {\bibfnamefont {C.-L.}\ \bibnamefont {Song}}, \bibinfo {author} {\bibfnamefont {X.}~\bibnamefont {Chen}}, \bibinfo {author} {\bibfnamefont {X.}~\bibnamefont {Ma}}, \bibinfo {author} {\bibfnamefont {Q.-k.}\ \bibnamefont {Xue}},\ and\ \bibinfo {author} {\bibfnamefont {C.}~\bibnamefont {Wu}},\ }\bibfield  {title} {\bibinfo {title} {Anisotropic vortex lattice structures in the fese superconductor},\ }\href {https://doi.org/10.1103/PhysRevB.85.104510} {\bibfield  {journal} {\bibinfo  {journal} {Physical Review B—Condensed Matter and Materials Physics}\ }\textbf {\bibinfo {volume} {85}},\ \bibinfo {pages} {104510} (\bibinfo {year} {2012})}\BibitemShut {NoStop}%
\bibitem [{\citenamefont {Greiter}(2011)}]{Greiter:2011}%
  \BibitemOpen
  \bibfield  {author} {\bibinfo {author} {\bibfnamefont {M.}~\bibnamefont {Greiter}},\ }\bibfield  {title} {\bibinfo {title} {Landau level quantization on the sphere},\ }\href {https://doi.org/10.1103/PhysRevB.83.115129} {\bibfield  {journal} {\bibinfo  {journal} {Physical Review B}\ }\textbf {\bibinfo {volume} {83}},\ \bibinfo {pages} {115129} (\bibinfo {year} {2011})}\BibitemShut {NoStop}%
\bibitem [{\citenamefont {Chang}\ \emph {et~al.}(2005)\citenamefont {Chang}, \citenamefont {Regnault}, \citenamefont {Jolicoeur},\ and\ \citenamefont {Jain}}]{Chang:2005}%
  \BibitemOpen
  \bibfield  {author} {\bibinfo {author} {\bibfnamefont {C.-C.}\ \bibnamefont {Chang}}, \bibinfo {author} {\bibfnamefont {N.}~\bibnamefont {Regnault}}, \bibinfo {author} {\bibfnamefont {T.}~\bibnamefont {Jolicoeur}},\ and\ \bibinfo {author} {\bibfnamefont {J.~K.}\ \bibnamefont {Jain}},\ }\bibfield  {title} {\bibinfo {title} {Composite fermionization of bosons in rapidly rotating atomic traps},\ }\href {https://doi.org/10.1103/PhysRevA.72.013611} {\bibfield  {journal} {\bibinfo  {journal} {Physical Review A—Atomic, Molecular, and Optical Physics}\ }\textbf {\bibinfo {volume} {72}},\ \bibinfo {pages} {013611} (\bibinfo {year} {2005})}\BibitemShut {NoStop}%
\bibitem [{\citenamefont {Tsui}\ \emph {et~al.}(1982{\natexlab{a}})\citenamefont {Tsui}, \citenamefont {Stormer},\ and\ \citenamefont {Gossard}}]{Tsui_PRL:1982}%
  \BibitemOpen
  \bibfield  {author} {\bibinfo {author} {\bibfnamefont {D.~C.}\ \bibnamefont {Tsui}}, \bibinfo {author} {\bibfnamefont {H.~L.}\ \bibnamefont {Stormer}},\ and\ \bibinfo {author} {\bibfnamefont {A.~C.}\ \bibnamefont {Gossard}},\ }\bibfield  {title} {\bibinfo {title} {Two-dimensional magnetotransport in the extreme quantum limit},\ }\href {https://doi.org/10.1103/PhysRevLett.48.1559} {\bibfield  {journal} {\bibinfo  {journal} {Physical Review Letters}\ }\textbf {\bibinfo {volume} {48}},\ \bibinfo {pages} {1559} (\bibinfo {year} {1982}{\natexlab{a}})}\BibitemShut {NoStop}%
\bibitem [{\citenamefont {Tsui}\ \emph {et~al.}(1982{\natexlab{b}})\citenamefont {Tsui}, \citenamefont {St{\"o}rmer},\ and\ \citenamefont {Gossard}}]{Tsui_PRB:1982}%
  \BibitemOpen
  \bibfield  {author} {\bibinfo {author} {\bibfnamefont {D.}~\bibnamefont {Tsui}}, \bibinfo {author} {\bibfnamefont {H.}~\bibnamefont {St{\"o}rmer}},\ and\ \bibinfo {author} {\bibfnamefont {A.}~\bibnamefont {Gossard}},\ }\bibfield  {title} {\bibinfo {title} {Zero-resistance state of two-dimensional electrons in a quantizing magnetic field},\ }\href {https://doi.org/10.1103/PhysRevB.25.1405} {\bibfield  {journal} {\bibinfo  {journal} {Physical Review B}\ }\textbf {\bibinfo {volume} {25}},\ \bibinfo {pages} {1405} (\bibinfo {year} {1982}{\natexlab{b}})}\BibitemShut {NoStop}%
\bibitem [{\citenamefont {Stormer}\ \emph {et~al.}(1983)\citenamefont {Stormer}, \citenamefont {Chang}, \citenamefont {Tsui}, \citenamefont {Hwang}, \citenamefont {Gossard},\ and\ \citenamefont {Wiegmann}}]{Stormer:1983}%
  \BibitemOpen
  \bibfield  {author} {\bibinfo {author} {\bibfnamefont {H.}~\bibnamefont {Stormer}}, \bibinfo {author} {\bibfnamefont {A.}~\bibnamefont {Chang}}, \bibinfo {author} {\bibfnamefont {D.}~\bibnamefont {Tsui}}, \bibinfo {author} {\bibfnamefont {J.}~\bibnamefont {Hwang}}, \bibinfo {author} {\bibfnamefont {A.}~\bibnamefont {Gossard}},\ and\ \bibinfo {author} {\bibfnamefont {W.}~\bibnamefont {Wiegmann}},\ }\bibfield  {title} {\bibinfo {title} {Fractional quantization of the hall effect},\ }\href {https://doi.org/10.1103/PhysRevLett.50.1953} {\bibfield  {journal} {\bibinfo  {journal} {Physical review letters}\ }\textbf {\bibinfo {volume} {50}},\ \bibinfo {pages} {1953} (\bibinfo {year} {1983})}\BibitemShut {NoStop}%
\bibitem [{\citenamefont {Prange}\ and\ \citenamefont {Girvin}(1990)}]{Prange:1990}%
  \BibitemOpen
  \bibfield  {author} {\bibinfo {author} {\bibfnamefont {R.}~\bibnamefont {Prange}}\ and\ \bibinfo {author} {\bibfnamefont {S.}~\bibnamefont {Girvin}},\ }\bibfield  {title} {\bibinfo {title} {The quantum hall effect},\ }\bibfield  {journal} {\bibinfo  {journal} {New York}\ }\href {https://doi.org/10.1007/978-1-4612-3350-3} {10.1007/978-1-4612-3350-3} (\bibinfo {year} {1990})\BibitemShut {NoStop}%
\bibitem [{\citenamefont {Stormer}(1992)}]{Stormer:1992}%
  \BibitemOpen
  \bibfield  {author} {\bibinfo {author} {\bibfnamefont {H.~L.}\ \bibnamefont {Stormer}},\ }\bibfield  {title} {\bibinfo {title} {Two-dimensional electron correlation in high magnetic fields},\ }\href {https://doi.org/10.1016/0921-4526(92)90138-I} {\bibfield  {journal} {\bibinfo  {journal} {Physica B: Condensed Matter}\ }\textbf {\bibinfo {volume} {177}},\ \bibinfo {pages} {401} (\bibinfo {year} {1992})}\BibitemShut {NoStop}%
\bibitem [{\citenamefont {Struck}\ and\ \citenamefont {Kramer}(2006)}]{Struck:2006}%
  \BibitemOpen
  \bibfield  {author} {\bibinfo {author} {\bibfnamefont {A.}~\bibnamefont {Struck}}\ and\ \bibinfo {author} {\bibfnamefont {B.}~\bibnamefont {Kramer}},\ }\bibfield  {title} {\bibinfo {title} {Electron correlations and single-particle physics in the integer quantum hall effect},\ }\href {https://doi.org/10.1103/PhysRevLett.97.106801} {\bibfield  {journal} {\bibinfo  {journal} {Physical review letters}\ }\textbf {\bibinfo {volume} {97}},\ \bibinfo {pages} {106801} (\bibinfo {year} {2006})}\BibitemShut {NoStop}%
\bibitem [{\citenamefont {Laughlin}(1983)}]{Laughlin:1983}%
  \BibitemOpen
  \bibfield  {author} {\bibinfo {author} {\bibfnamefont {R.~B.}\ \bibnamefont {Laughlin}},\ }\bibfield  {title} {\bibinfo {title} {Anomalous quantum hall effect: an incompressible quantum fluid with fractionally charged excitations},\ }\href {https://doi.org/10.1103/PhysRevLett.50.1395} {\bibfield  {journal} {\bibinfo  {journal} {Physical Review Letters}\ }\textbf {\bibinfo {volume} {50}},\ \bibinfo {pages} {1395} (\bibinfo {year} {1983})}\BibitemShut {NoStop}%
\bibitem [{\citenamefont {Shi}\ \emph {et~al.}(2020)\citenamefont {Shi}, \citenamefont {Shih}, \citenamefont {Gustafsson}, \citenamefont {Rhodes}, \citenamefont {Kim}, \citenamefont {Watanabe}, \citenamefont {Taniguchi}, \citenamefont {Papi{\'c}}, \citenamefont {Hone},\ and\ \citenamefont {Dean}}]{Shi:2020}%
  \BibitemOpen
  \bibfield  {author} {\bibinfo {author} {\bibfnamefont {Q.}~\bibnamefont {Shi}}, \bibinfo {author} {\bibfnamefont {E.-M.}\ \bibnamefont {Shih}}, \bibinfo {author} {\bibfnamefont {M.~V.}\ \bibnamefont {Gustafsson}}, \bibinfo {author} {\bibfnamefont {D.~A.}\ \bibnamefont {Rhodes}}, \bibinfo {author} {\bibfnamefont {B.}~\bibnamefont {Kim}}, \bibinfo {author} {\bibfnamefont {K.}~\bibnamefont {Watanabe}}, \bibinfo {author} {\bibfnamefont {T.}~\bibnamefont {Taniguchi}}, \bibinfo {author} {\bibfnamefont {Z.}~\bibnamefont {Papi{\'c}}}, \bibinfo {author} {\bibfnamefont {J.}~\bibnamefont {Hone}},\ and\ \bibinfo {author} {\bibfnamefont {C.~R.}\ \bibnamefont {Dean}},\ }\bibfield  {title} {\bibinfo {title} {Odd-and even-denominator fractional quantum hall states in monolayer wse2},\ }\href {https://doi.org/10.1038/s41565-020-0685-6} {\bibfield  {journal} {\bibinfo  {journal} {Nature Nanotechnology}\ }\textbf {\bibinfo {volume} {15}},\ \bibinfo {pages} {569} (\bibinfo {year} {2020})}\BibitemShut {NoStop}%
\bibitem [{\citenamefont {Chen}\ \emph {et~al.}(2024)\citenamefont {Chen}, \citenamefont {Huang}, \citenamefont {Li}, \citenamefont {Tong}, \citenamefont {Kuang}, \citenamefont {Xi}, \citenamefont {Watanabe}, \citenamefont {Taniguchi}, \citenamefont {Liu}, \citenamefont {Zhu} \emph {et~al.}}]{Chen:2024}%
  \BibitemOpen
  \bibfield  {author} {\bibinfo {author} {\bibfnamefont {Y.}~\bibnamefont {Chen}}, \bibinfo {author} {\bibfnamefont {Y.}~\bibnamefont {Huang}}, \bibinfo {author} {\bibfnamefont {Q.}~\bibnamefont {Li}}, \bibinfo {author} {\bibfnamefont {B.}~\bibnamefont {Tong}}, \bibinfo {author} {\bibfnamefont {G.}~\bibnamefont {Kuang}}, \bibinfo {author} {\bibfnamefont {C.}~\bibnamefont {Xi}}, \bibinfo {author} {\bibfnamefont {K.}~\bibnamefont {Watanabe}}, \bibinfo {author} {\bibfnamefont {T.}~\bibnamefont {Taniguchi}}, \bibinfo {author} {\bibfnamefont {G.}~\bibnamefont {Liu}}, \bibinfo {author} {\bibfnamefont {Z.}~\bibnamefont {Zhu}}, \emph {et~al.},\ }\bibfield  {title} {\bibinfo {title} {Tunable even-and odd-denominator fractional quantum hall states in trilayer graphene},\ }\href {https://doi.org/10.1038/s41467-024-50589-2} {\bibfield  {journal} {\bibinfo  {journal} {Nature Communications}\ }\textbf {\bibinfo {volume} {15}},\ \bibinfo {pages} {6236} (\bibinfo {year} {2024})}\BibitemShut {NoStop}%
\bibitem [{\citenamefont {Feldman}\ and\ \citenamefont {Halperin}(2021)}]{Feldman:2021}%
  \BibitemOpen
  \bibfield  {author} {\bibinfo {author} {\bibfnamefont {D.~E.}\ \bibnamefont {Feldman}}\ and\ \bibinfo {author} {\bibfnamefont {B.~I.}\ \bibnamefont {Halperin}},\ }\bibfield  {title} {\bibinfo {title} {Fractional charge and fractional statistics in the quantum hall effects},\ }\href {https://doi.org/10.1088/1361-6633/ac03aa} {\bibfield  {journal} {\bibinfo  {journal} {Reports on Progress in Physics}\ }\textbf {\bibinfo {volume} {84}},\ \bibinfo {pages} {076501} (\bibinfo {year} {2021})}\BibitemShut {NoStop}%
\bibitem [{\citenamefont {Bruno}\ \emph {et~al.}(2012)\citenamefont {Bruno}, \citenamefont {Dugaev},\ and\ \citenamefont {Taillefumier}}]{Bruno:2012}%
  \BibitemOpen
  \bibfield  {author} {\bibinfo {author} {\bibfnamefont {P.}~\bibnamefont {Bruno}}, \bibinfo {author} {\bibfnamefont {V.}~\bibnamefont {Dugaev}},\ and\ \bibinfo {author} {\bibfnamefont {M.}~\bibnamefont {Taillefumier}},\ }\bibfield  {title} {\bibinfo {title} {Topological hall effect in magnetic nanostructures},\ }\href {https://doi.org/https://www-old.mpi-halle.mpg.de/annual_reports/2004.pdf/jb_results_24.pdf} {\bibfield  {journal} {\bibinfo  {journal} {Phys. Rev. Lett}\ }\textbf {\bibinfo {volume} {93}},\ \bibinfo {pages} {58} (\bibinfo {year} {2012})}\BibitemShut {NoStop}%
\bibitem [{\citenamefont {Sprinkart}\ \emph {et~al.}(2024)\citenamefont {Sprinkart}, \citenamefont {Scheer},\ and\ \citenamefont {Di~Bernardo}}]{Sprinkart:2024}%
  \BibitemOpen
  \bibfield  {author} {\bibinfo {author} {\bibfnamefont {N.}~\bibnamefont {Sprinkart}}, \bibinfo {author} {\bibfnamefont {E.}~\bibnamefont {Scheer}},\ and\ \bibinfo {author} {\bibfnamefont {A.}~\bibnamefont {Di~Bernardo}},\ }\bibfield  {title} {\bibinfo {title} {Tutorial: From topology to hall effects—implications of berry phase physics},\ }\href {https://doi.org/10.1007/s10909-024-03219-6} {\bibfield  {journal} {\bibinfo  {journal} {Journal of Low Temperature Physics}\ ,\ \bibinfo {pages} {1}} (\bibinfo {year} {2024})}\BibitemShut {NoStop}%
\bibitem [{\citenamefont {Jain}(1990)}]{Jain:1990}%
  \BibitemOpen
  \bibfield  {author} {\bibinfo {author} {\bibfnamefont {J.}~\bibnamefont {Jain}},\ }\bibfield  {title} {\bibinfo {title} {Theory of the fractional quantum hall effect},\ }\href {https://doi.org/10.1103/PhysRevB.41.7653} {\bibfield  {journal} {\bibinfo  {journal} {Physical Review B}\ }\textbf {\bibinfo {volume} {41}},\ \bibinfo {pages} {7653} (\bibinfo {year} {1990})}\BibitemShut {NoStop}%
\bibitem [{\citenamefont {Cooper}\ and\ \citenamefont {Wilkin}(1999)}]{Cooper:1999}%
  \BibitemOpen
  \bibfield  {author} {\bibinfo {author} {\bibfnamefont {N.}~\bibnamefont {Cooper}}\ and\ \bibinfo {author} {\bibfnamefont {N.}~\bibnamefont {Wilkin}},\ }\bibfield  {title} {\bibinfo {title} {Composite fermion description of rotating bose-einstein condensates},\ }\href {https://doi.org/10.1103/PhysRevB.60.R16279} {\bibfield  {journal} {\bibinfo  {journal} {Physical Review B}\ }\textbf {\bibinfo {volume} {60}},\ \bibinfo {pages} {R16279} (\bibinfo {year} {1999})}\BibitemShut {NoStop}%
\bibitem [{\citenamefont {Leinaas}\ and\ \citenamefont {Myrheim}(1977)}]{Leinaas:1977}%
  \BibitemOpen
  \bibfield  {author} {\bibinfo {author} {\bibfnamefont {J.}~\bibnamefont {Leinaas}}\ and\ \bibinfo {author} {\bibfnamefont {J.}~\bibnamefont {Myrheim}},\ }\bibfield  {title} {\bibinfo {title} {On the theory of identical particles},\ }\href {https://doi.org/10.1007/BF02727953} {\bibfield  {journal} {\bibinfo  {journal} {Il nuovo cimento}\ }\textbf {\bibinfo {volume} {37}},\ \bibinfo {pages} {132} (\bibinfo {year} {1977})}\BibitemShut {NoStop}%
\bibitem [{\citenamefont {Wilczek}(1982)}]{Wilczek:1982}%
  \BibitemOpen
  \bibfield  {author} {\bibinfo {author} {\bibfnamefont {F.}~\bibnamefont {Wilczek}},\ }\bibfield  {title} {\bibinfo {title} {Magnetic flux, angular momentum, and statistics},\ }\href {https://doi.org/10.1103/PhysRevLett.48.1144} {\bibfield  {journal} {\bibinfo  {journal} {Physical Review Letters}\ }\textbf {\bibinfo {volume} {48}},\ \bibinfo {pages} {1144} (\bibinfo {year} {1982})}\BibitemShut {NoStop}%
\bibitem [{\citenamefont {Jain}(1989)}]{Jain:1989}%
  \BibitemOpen
  \bibfield  {author} {\bibinfo {author} {\bibfnamefont {J.~K.}\ \bibnamefont {Jain}},\ }\bibfield  {title} {\bibinfo {title} {Composite-fermion approach for the fractional quantum hall effect},\ }\href {https://doi.org/10.1103/PhysRevLett.63.199} {\bibfield  {journal} {\bibinfo  {journal} {Physical review letters}\ }\textbf {\bibinfo {volume} {63}},\ \bibinfo {pages} {199} (\bibinfo {year} {1989})}\BibitemShut {NoStop}%
\bibitem [{\citenamefont {Shankar}\ and\ \citenamefont {Murthy}(1997)}]{Shankar:1997}%
  \BibitemOpen
  \bibfield  {author} {\bibinfo {author} {\bibfnamefont {R.}~\bibnamefont {Shankar}}\ and\ \bibinfo {author} {\bibfnamefont {G.}~\bibnamefont {Murthy}},\ }\bibfield  {title} {\bibinfo {title} {Towards a field theory of fractional quantum hall states},\ }\href {https://doi.org/10.1103/PhysRevLett.79.4437} {\bibfield  {journal} {\bibinfo  {journal} {Physical review letters}\ }\textbf {\bibinfo {volume} {79}},\ \bibinfo {pages} {4437} (\bibinfo {year} {1997})}\BibitemShut {NoStop}%
\bibitem [{\citenamefont {Lopez}\ and\ \citenamefont {Fradkin}(1991)}]{Lopez:1991}%
  \BibitemOpen
  \bibfield  {author} {\bibinfo {author} {\bibfnamefont {A.}~\bibnamefont {Lopez}}\ and\ \bibinfo {author} {\bibfnamefont {E.}~\bibnamefont {Fradkin}},\ }\bibfield  {title} {\bibinfo {title} {Fractional quantum hall effect and chern-simons gauge theories},\ }\href {https://doi.org/10.1103/PhysRevB.44.5246} {\bibfield  {journal} {\bibinfo  {journal} {Physical Review B}\ }\textbf {\bibinfo {volume} {44}},\ \bibinfo {pages} {5246} (\bibinfo {year} {1991})}\BibitemShut {NoStop}%
\bibitem [{\citenamefont {Modak}\ \emph {et~al.}(2011)\citenamefont {Modak}, \citenamefont {Mandal},\ and\ \citenamefont {Sengupta}}]{Modak:2011}%
  \BibitemOpen
  \bibfield  {author} {\bibinfo {author} {\bibfnamefont {S.}~\bibnamefont {Modak}}, \bibinfo {author} {\bibfnamefont {S.~S.}\ \bibnamefont {Mandal}},\ and\ \bibinfo {author} {\bibfnamefont {K.}~\bibnamefont {Sengupta}},\ }\bibfield  {title} {\bibinfo {title} {Fermionic chern-simons theory of su (4) fractional quantum hall effect},\ }\href {https://doi.org/10.1103/PhysRevB.84.165118} {\bibfield  {journal} {\bibinfo  {journal} {Physical Review B—Condensed Matter and Materials Physics}\ }\textbf {\bibinfo {volume} {84}},\ \bibinfo {pages} {165118} (\bibinfo {year} {2011})}\BibitemShut {NoStop}%
\bibitem [{\citenamefont {Miesner}\ \emph {et~al.}(1999)\citenamefont {Miesner}, \citenamefont {Stamper-Kurn}, \citenamefont {Stenger}, \citenamefont {Inouye}, \citenamefont {Chikkatur},\ and\ \citenamefont {Ketterle}}]{Miesner:1999}%
  \BibitemOpen
  \bibfield  {author} {\bibinfo {author} {\bibfnamefont {H.-J.}\ \bibnamefont {Miesner}}, \bibinfo {author} {\bibfnamefont {D.}~\bibnamefont {Stamper-Kurn}}, \bibinfo {author} {\bibfnamefont {J.}~\bibnamefont {Stenger}}, \bibinfo {author} {\bibfnamefont {S.}~\bibnamefont {Inouye}}, \bibinfo {author} {\bibfnamefont {A.}~\bibnamefont {Chikkatur}},\ and\ \bibinfo {author} {\bibfnamefont {W.}~\bibnamefont {Ketterle}},\ }\bibfield  {title} {\bibinfo {title} {Observation of metastable states in spinor bose-einstein condensates},\ }\href {https://doi.org/10.1103/PhysRevLett.82.2228} {\bibfield  {journal} {\bibinfo  {journal} {Physical Review Letters}\ }\textbf {\bibinfo {volume} {82}},\ \bibinfo {pages} {2228} (\bibinfo {year} {1999})}\BibitemShut {NoStop}%
\bibitem [{\citenamefont {Modugno}\ \emph {et~al.}(2002)\citenamefont {Modugno}, \citenamefont {Modugno}, \citenamefont {Riboli}, \citenamefont {Roati},\ and\ \citenamefont {Inguscio}}]{Modugno:2002}%
  \BibitemOpen
  \bibfield  {author} {\bibinfo {author} {\bibfnamefont {G.}~\bibnamefont {Modugno}}, \bibinfo {author} {\bibfnamefont {M.}~\bibnamefont {Modugno}}, \bibinfo {author} {\bibfnamefont {F.}~\bibnamefont {Riboli}}, \bibinfo {author} {\bibfnamefont {G.}~\bibnamefont {Roati}},\ and\ \bibinfo {author} {\bibfnamefont {M.}~\bibnamefont {Inguscio}},\ }\bibfield  {title} {\bibinfo {title} {Two atomic species superfluid},\ }\href {https://doi.org/10.1103/PhysRevLett.89.190404} {\bibfield  {journal} {\bibinfo  {journal} {Physical Review Letters}\ }\textbf {\bibinfo {volume} {89}},\ \bibinfo {pages} {190404} (\bibinfo {year} {2002})}\BibitemShut {NoStop}%
\bibitem [{\citenamefont {Thalhammer}\ \emph {et~al.}(2008)\citenamefont {Thalhammer}, \citenamefont {Barontini}, \citenamefont {De~Sarlo}, \citenamefont {Catani}, \citenamefont {Minardi},\ and\ \citenamefont {Inguscio}}]{Thalhammer:2008}%
  \BibitemOpen
  \bibfield  {author} {\bibinfo {author} {\bibfnamefont {G.}~\bibnamefont {Thalhammer}}, \bibinfo {author} {\bibfnamefont {G.}~\bibnamefont {Barontini}}, \bibinfo {author} {\bibfnamefont {L.}~\bibnamefont {De~Sarlo}}, \bibinfo {author} {\bibfnamefont {J.}~\bibnamefont {Catani}}, \bibinfo {author} {\bibfnamefont {F.}~\bibnamefont {Minardi}},\ and\ \bibinfo {author} {\bibfnamefont {M.}~\bibnamefont {Inguscio}},\ }\bibfield  {title} {\bibinfo {title} {Double species bose-einstein condensate with tunable interspecies interactions},\ }\href {https://doi.org/10.1103/PhysRevLett.100.210402} {\bibfield  {journal} {\bibinfo  {journal} {Physical review letters}\ }\textbf {\bibinfo {volume} {100}},\ \bibinfo {pages} {210402} (\bibinfo {year} {2008})}\BibitemShut {NoStop}%
\bibitem [{\citenamefont {Papp}\ \emph {et~al.}(2008)\citenamefont {Papp}, \citenamefont {Pino},\ and\ \citenamefont {Wieman}}]{Papp:2008}%
  \BibitemOpen
  \bibfield  {author} {\bibinfo {author} {\bibfnamefont {S.}~\bibnamefont {Papp}}, \bibinfo {author} {\bibfnamefont {J.}~\bibnamefont {Pino}},\ and\ \bibinfo {author} {\bibfnamefont {C.}~\bibnamefont {Wieman}},\ }\bibfield  {title} {\bibinfo {title} {Tunable miscibility in a dual-species bose-einstein condensate},\ }\href {https://doi.org/10.1103/PhysRevLett.101.040402} {\bibfield  {journal} {\bibinfo  {journal} {Physical review letters}\ }\textbf {\bibinfo {volume} {101}},\ \bibinfo {pages} {040402} (\bibinfo {year} {2008})}\BibitemShut {NoStop}%
\bibitem [{\citenamefont {Kukushkin}\ \emph {et~al.}(1999)\citenamefont {Kukushkin}, \citenamefont {Klitzing},\ and\ \citenamefont {Eberl}}]{Kukushkin:1999}%
  \BibitemOpen
  \bibfield  {author} {\bibinfo {author} {\bibfnamefont {I.}~\bibnamefont {Kukushkin}}, \bibinfo {author} {\bibfnamefont {K.~v.}\ \bibnamefont {Klitzing}},\ and\ \bibinfo {author} {\bibfnamefont {K.}~\bibnamefont {Eberl}},\ }\bibfield  {title} {\bibinfo {title} {Spin polarization of composite fermions: Measurements of the fermi energy},\ }\href {https://doi.org/10.1103/PhysRevLett.82.3665} {\bibfield  {journal} {\bibinfo  {journal} {Physical review letters}\ }\textbf {\bibinfo {volume} {82}},\ \bibinfo {pages} {3665} (\bibinfo {year} {1999})}\BibitemShut {NoStop}%
\bibitem [{\citenamefont {Indra}\ \emph {et~al.}(2018)\citenamefont {Indra}, \citenamefont {Das},\ and\ \citenamefont {Majumder}}]{Indra:2018}%
  \BibitemOpen
  \bibfield  {author} {\bibinfo {author} {\bibfnamefont {M.}~\bibnamefont {Indra}}, \bibinfo {author} {\bibfnamefont {D.}~\bibnamefont {Das}},\ and\ \bibinfo {author} {\bibfnamefont {D.}~\bibnamefont {Majumder}},\ }\bibfield  {title} {\bibinfo {title} {Study of partially polarized fractional quantum hall states},\ }\href {https://doi.org/10.1016/j.physleta.2018.08.008} {\bibfield  {journal} {\bibinfo  {journal} {Physics Letters A}\ }\textbf {\bibinfo {volume} {382}},\ \bibinfo {pages} {2984} (\bibinfo {year} {2018})}\BibitemShut {NoStop}%
\bibitem [{\citenamefont {Balram}\ \emph {et~al.}(2015)\citenamefont {Balram}, \citenamefont {T{\H{o}}ke}, \citenamefont {W{\'o}js},\ and\ \citenamefont {Jain}}]{Balram:2015}%
  \BibitemOpen
  \bibfield  {author} {\bibinfo {author} {\bibfnamefont {A.~C.}\ \bibnamefont {Balram}}, \bibinfo {author} {\bibfnamefont {C.}~\bibnamefont {T{\H{o}}ke}}, \bibinfo {author} {\bibfnamefont {A.}~\bibnamefont {W{\'o}js}},\ and\ \bibinfo {author} {\bibfnamefont {J.~K.}\ \bibnamefont {Jain}},\ }\bibfield  {title} {\bibinfo {title} {Phase diagram of fractional quantum hall effect of composite fermions in multicomponent systems},\ }\href {https://doi.org/10.1103/PhysRevB.91.045109} {\bibfield  {journal} {\bibinfo  {journal} {Physical review B}\ }\textbf {\bibinfo {volume} {91}},\ \bibinfo {pages} {045109} (\bibinfo {year} {2015})}\BibitemShut {NoStop}%
\bibitem [{\citenamefont {Abo-Shaeer}\ \emph {et~al.}(2001)\citenamefont {Abo-Shaeer}, \citenamefont {Raman}, \citenamefont {Vogels},\ and\ \citenamefont {Ketterle}}]{Abo:2001}%
  \BibitemOpen
  \bibfield  {author} {\bibinfo {author} {\bibfnamefont {J.~R.}\ \bibnamefont {Abo-Shaeer}}, \bibinfo {author} {\bibfnamefont {C.}~\bibnamefont {Raman}}, \bibinfo {author} {\bibfnamefont {J.~M.}\ \bibnamefont {Vogels}},\ and\ \bibinfo {author} {\bibfnamefont {W.}~\bibnamefont {Ketterle}},\ }\bibfield  {title} {\bibinfo {title} {Observation of vortex lattices in bose-einstein condensates},\ }\href {https://doi.org/10.1126/science.1060182} {\bibfield  {journal} {\bibinfo  {journal} {Science}\ }\textbf {\bibinfo {volume} {292}},\ \bibinfo {pages} {476} (\bibinfo {year} {2001})}\BibitemShut {NoStop}%
\bibitem [{\citenamefont {Fetter}(2009)}]{Fetter:2009}%
  \BibitemOpen
  \bibfield  {author} {\bibinfo {author} {\bibfnamefont {A.~L.}\ \bibnamefont {Fetter}},\ }\bibfield  {title} {\bibinfo {title} {Rotating trapped bose-einstein condensates},\ }\href {https://doi.org/10.1103/RevModPhys.81.647} {\bibfield  {journal} {\bibinfo  {journal} {Reviews of Modern Physics}\ }\textbf {\bibinfo {volume} {81}},\ \bibinfo {pages} {647} (\bibinfo {year} {2009})}\BibitemShut {NoStop}%
\bibitem [{\citenamefont {Baranov}\ \emph {et~al.}(2005)\citenamefont {Baranov}, \citenamefont {Osterloh},\ and\ \citenamefont {Lewenstein}}]{Baranov:2005}%
  \BibitemOpen
  \bibfield  {author} {\bibinfo {author} {\bibfnamefont {M.}~\bibnamefont {Baranov}}, \bibinfo {author} {\bibfnamefont {K.}~\bibnamefont {Osterloh}},\ and\ \bibinfo {author} {\bibfnamefont {M.}~\bibnamefont {Lewenstein}},\ }\bibfield  {title} {\bibinfo {title} {Fractional quantum hall states in ultracold rapidly rotating dipolar fermi gases},\ }\href {https://doi.org/10.1103/PhysRevLett.94.070404} {\bibfield  {journal} {\bibinfo  {journal} {Physical review letters}\ }\textbf {\bibinfo {volume} {94}},\ \bibinfo {pages} {070404} (\bibinfo {year} {2005})}\BibitemShut {NoStop}%
\bibitem [{\citenamefont {Pino}\ \emph {et~al.}(2013)\citenamefont {Pino}, \citenamefont {Alba}, \citenamefont {Taron}, \citenamefont {Garc{\'\i}a-Ripoll},\ and\ \citenamefont {Barber{\'a}n}}]{Pino:2013}%
  \BibitemOpen
  \bibfield  {author} {\bibinfo {author} {\bibfnamefont {H.}~\bibnamefont {Pino}}, \bibinfo {author} {\bibfnamefont {E.}~\bibnamefont {Alba}}, \bibinfo {author} {\bibfnamefont {J.}~\bibnamefont {Taron}}, \bibinfo {author} {\bibfnamefont {J.~J.}\ \bibnamefont {Garc{\'\i}a-Ripoll}},\ and\ \bibinfo {author} {\bibfnamefont {N.}~\bibnamefont {Barber{\'a}n}},\ }\bibfield  {title} {\bibinfo {title} {Hall response of interacting bosonic atoms in strong gauge fields: From condensed to fractional-quantum-hall states},\ }\href {https://doi.org/10.1103/PhysRevA.87.053611} {\bibfield  {journal} {\bibinfo  {journal} {Physical Review A—Atomic, Molecular, and Optical Physics}\ }\textbf {\bibinfo {volume} {87}},\ \bibinfo {pages} {053611} (\bibinfo {year} {2013})}\BibitemShut {NoStop}%
\bibitem [{\citenamefont {Imran}\ and\ \citenamefont {Ahsan}(2023)}]{Imran:2023}%
  \BibitemOpen
  \bibfield  {author} {\bibinfo {author} {\bibfnamefont {M.}~\bibnamefont {Imran}}\ and\ \bibinfo {author} {\bibfnamefont {M.}~\bibnamefont {Ahsan}},\ }\bibfield  {title} {\bibinfo {title} {Novel phases in rotating bose-condensed gas: vortices and quantum correlation},\ }\href {https://doi.org/10.1140/epjd/s10053-023-00662-0} {\bibfield  {journal} {\bibinfo  {journal} {The European Physical Journal D}\ }\textbf {\bibinfo {volume} {77}},\ \bibinfo {pages} {79} (\bibinfo {year} {2023})}\BibitemShut {NoStop}%
\bibitem [{\citenamefont {Das}\ \emph {et~al.}(2018)\citenamefont {Das}, \citenamefont {Sahu},\ and\ \citenamefont {Majumder}}]{Das:2018}%
  \BibitemOpen
  \bibfield  {author} {\bibinfo {author} {\bibfnamefont {D.}~\bibnamefont {Das}}, \bibinfo {author} {\bibfnamefont {S.}~\bibnamefont {Sahu}},\ and\ \bibinfo {author} {\bibfnamefont {D.}~\bibnamefont {Majumder}},\ }\bibfield  {title} {\bibinfo {title} {Roton minimum at $\nu$ = 1/ 2 filled fractional quantum hall effect of bose particles},\ }\href {https://doi.org/10.1016/j.physb.2018.08.042} {\bibfield  {journal} {\bibinfo  {journal} {Physica B: Condensed Matter}\ }\textbf {\bibinfo {volume} {550}},\ \bibinfo {pages} {96} (\bibinfo {year} {2018})}\BibitemShut {NoStop}%
\bibitem [{\citenamefont {Indra}\ and\ \citenamefont {Majumder}(2020)}]{Indra:2020}%
  \BibitemOpen
  \bibfield  {author} {\bibinfo {author} {\bibfnamefont {M.}~\bibnamefont {Indra}}\ and\ \bibinfo {author} {\bibfnamefont {D.}~\bibnamefont {Majumder}},\ }\bibfield  {title} {\bibinfo {title} {Collective spin density excitation of fractional quantum hall states in dilute ultra-cold bose atoms},\ }\href {https://doi.org/10.1016/j.ssc.2019.113796} {\bibfield  {journal} {\bibinfo  {journal} {Solid State Communications}\ }\textbf {\bibinfo {volume} {306}},\ \bibinfo {pages} {113796} (\bibinfo {year} {2020})}\BibitemShut {NoStop}%
\bibitem [{\citenamefont {Indra}\ \emph {et~al.}(2023)\citenamefont {Indra}, \citenamefont {Jain},\ and\ \citenamefont {Mondal}}]{Indra:2023}%
  \BibitemOpen
  \bibfield  {author} {\bibinfo {author} {\bibfnamefont {M.}~\bibnamefont {Indra}}, \bibinfo {author} {\bibfnamefont {D.}~\bibnamefont {Jain}},\ and\ \bibinfo {author} {\bibfnamefont {S.}~\bibnamefont {Mondal}},\ }\bibfield  {title} {\bibinfo {title} {Double roton-minima in bosonic fractional quantum hall states},\ }\href {https://doi.org/10.1088/1402-4896/acd426} {\bibfield  {journal} {\bibinfo  {journal} {Physica Scripta}\ }\textbf {\bibinfo {volume} {98}},\ \bibinfo {pages} {065948} (\bibinfo {year} {2023})}\BibitemShut {NoStop}%
\bibitem [{\citenamefont {Indra}\ and\ \citenamefont {Mondal}(2024)}]{Indra:2024}%
  \BibitemOpen
  \bibfield  {author} {\bibinfo {author} {\bibfnamefont {M.}~\bibnamefont {Indra}}\ and\ \bibinfo {author} {\bibfnamefont {S.}~\bibnamefont {Mondal}},\ }\bibfield  {title} {\bibinfo {title} {Collective excitation of bosonic quantum hall state},\ }\href {https://doi.org/10.1007/s10909-023-03023-8} {\bibfield  {journal} {\bibinfo  {journal} {Journal of Low Temperature Physics}\ }\textbf {\bibinfo {volume} {214}},\ \bibinfo {pages} {294} (\bibinfo {year} {2024})}\BibitemShut {NoStop}%
\bibitem [{\citenamefont {Wilkin}\ and\ \citenamefont {Gunn}(2000)}]{Wilkin:2000}%
  \BibitemOpen
  \bibfield  {author} {\bibinfo {author} {\bibfnamefont {N.}~\bibnamefont {Wilkin}}\ and\ \bibinfo {author} {\bibfnamefont {J.}~\bibnamefont {Gunn}},\ }\bibfield  {title} {\bibinfo {title} {Condensation of “composite bosons” in a rotating bec},\ }\href {https://doi.org/10.1103/PhysRevLett.84.6} {\bibfield  {journal} {\bibinfo  {journal} {Physical review letters}\ }\textbf {\bibinfo {volume} {84}},\ \bibinfo {pages} {6} (\bibinfo {year} {2000})}\BibitemShut {NoStop}%
\bibitem [{\citenamefont {Cooper}\ \emph {et~al.}(2001)\citenamefont {Cooper}, \citenamefont {Wilkin},\ and\ \citenamefont {Gunn}}]{Cooper:2001}%
  \BibitemOpen
  \bibfield  {author} {\bibinfo {author} {\bibfnamefont {N.~R.}\ \bibnamefont {Cooper}}, \bibinfo {author} {\bibfnamefont {N.~K.}\ \bibnamefont {Wilkin}},\ and\ \bibinfo {author} {\bibfnamefont {J.}~\bibnamefont {Gunn}},\ }\bibfield  {title} {\bibinfo {title} {Quantum phases of vortices in rotating bose-einstein condensates},\ }\href {https://doi.org/10.1103/PhysRevLett.87.120405} {\bibfield  {journal} {\bibinfo  {journal} {Physical review letters}\ }\textbf {\bibinfo {volume} {87}},\ \bibinfo {pages} {120405} (\bibinfo {year} {2001})}\BibitemShut {NoStop}%
\bibitem [{\citenamefont {Chung}\ and\ \citenamefont {Jolicoeur}(2008)}]{Chung:2008}%
  \BibitemOpen
  \bibfield  {author} {\bibinfo {author} {\bibfnamefont {B.}~\bibnamefont {Chung}}\ and\ \bibinfo {author} {\bibfnamefont {T.}~\bibnamefont {Jolicoeur}},\ }\bibfield  {title} {\bibinfo {title} {Fermions out of dipolar bosons in the lowest landau level},\ }\href {https://doi.org/10.1103/PhysRevA.77.043608} {\bibfield  {journal} {\bibinfo  {journal} {Physical Review A—Atomic, Molecular, and Optical Physics}\ }\textbf {\bibinfo {volume} {77}},\ \bibinfo {pages} {043608} (\bibinfo {year} {2008})}\BibitemShut {NoStop}%
\bibitem [{\citenamefont {Simon}\ \emph {et~al.}(2007)\citenamefont {Simon}, \citenamefont {Rezayi},\ and\ \citenamefont {Cooper}}]{Simon:2007}%
  \BibitemOpen
  \bibfield  {author} {\bibinfo {author} {\bibfnamefont {S.~H.}\ \bibnamefont {Simon}}, \bibinfo {author} {\bibfnamefont {E.}~\bibnamefont {Rezayi}},\ and\ \bibinfo {author} {\bibfnamefont {N.~R.}\ \bibnamefont {Cooper}},\ }\bibfield  {title} {\bibinfo {title} {Pseudopotentials for multiparticle interactions in the quantum hall regime},\ }\href {https://doi.org/10.1103/PhysRevB.75.195306} {\bibfield  {journal} {\bibinfo  {journal} {Physical Review B—Condensed Matter and Materials Physics}\ }\textbf {\bibinfo {volume} {75}},\ \bibinfo {pages} {195306} (\bibinfo {year} {2007})}\BibitemShut {NoStop}%
\bibitem [{\citenamefont {Gong}\ \emph {et~al.}(2014)\citenamefont {Gong}, \citenamefont {Zhu},\ and\ \citenamefont {Sheng}}]{Gong:2014}%
  \BibitemOpen
  \bibfield  {author} {\bibinfo {author} {\bibfnamefont {S.-S.}\ \bibnamefont {Gong}}, \bibinfo {author} {\bibfnamefont {W.}~\bibnamefont {Zhu}},\ and\ \bibinfo {author} {\bibfnamefont {D.}~\bibnamefont {Sheng}},\ }\bibfield  {title} {\bibinfo {title} {Emergent chiral spin liquid: Fractional quantum hall effect in a kagome heisenberg model},\ }\href {https://doi.org/10.1038/srep06317} {\bibfield  {journal} {\bibinfo  {journal} {Scientific reports}\ }\textbf {\bibinfo {volume} {4}},\ \bibinfo {pages} {6317} (\bibinfo {year} {2014})}\BibitemShut {NoStop}%
\bibitem [{\citenamefont {Majumder}\ \emph {et~al.}(2009)\citenamefont {Majumder}, \citenamefont {Mandal},\ and\ \citenamefont {Jain}}]{Majumder:2009}%
  \BibitemOpen
  \bibfield  {author} {\bibinfo {author} {\bibfnamefont {D.}~\bibnamefont {Majumder}}, \bibinfo {author} {\bibfnamefont {S.~S.}\ \bibnamefont {Mandal}},\ and\ \bibinfo {author} {\bibfnamefont {J.~K.}\ \bibnamefont {Jain}},\ }\bibfield  {title} {\bibinfo {title} {Collective excitations of composite fermions across multiple $\lambda$ levels},\ }\href {https://doi.org/10.1038/nphys1275} {\bibfield  {journal} {\bibinfo  {journal} {Nature Physics}\ }\textbf {\bibinfo {volume} {5}},\ \bibinfo {pages} {403} (\bibinfo {year} {2009})}\BibitemShut {NoStop}%
\bibitem [{\citenamefont {Haldane}(1983)}]{Haldane:1983}%
  \BibitemOpen
  \bibfield  {author} {\bibinfo {author} {\bibfnamefont {F.~D.~M.}\ \bibnamefont {Haldane}},\ }\bibfield  {title} {\bibinfo {title} {Fractional quantization of the hall effect: A hierarchy of incompressible quantum fluid states},\ }\href {https://doi.org/10.1103/PhysRevLett.51.605} {\bibfield  {journal} {\bibinfo  {journal} {Physical Review Letters}\ }\textbf {\bibinfo {volume} {51}},\ \bibinfo {pages} {605} (\bibinfo {year} {1983})}\BibitemShut {NoStop}%
\bibitem [{\citenamefont {Jain}(2007)}]{Jain:2007}%
  \BibitemOpen
  \bibfield  {author} {\bibinfo {author} {\bibfnamefont {J.~K.}\ \bibnamefont {Jain}},\ }\href@noop {} {\emph {\bibinfo {title} {Composite fermions}}}\ (\bibinfo  {publisher} {Cambridge University Press},\ \bibinfo {year} {2007})\BibitemShut {NoStop}%
\bibitem [{\citenamefont {Haldane}\ and\ \citenamefont {Rezayi}(1985)}]{Haldane:1985}%
  \BibitemOpen
  \bibfield  {author} {\bibinfo {author} {\bibfnamefont {F.~D.~M.}\ \bibnamefont {Haldane}}\ and\ \bibinfo {author} {\bibfnamefont {E.~H.}\ \bibnamefont {Rezayi}},\ }\bibfield  {title} {\bibinfo {title} {Finite-size studies of the incompressible state of the fractionally quantized hall effect and its excitations},\ }\href {https://doi.org/10.1103/PhysRevLett.54.237} {\bibfield  {journal} {\bibinfo  {journal} {Physical review letters}\ }\textbf {\bibinfo {volume} {54}},\ \bibinfo {pages} {237} (\bibinfo {year} {1985})}\BibitemShut {NoStop}%
\bibitem [{\citenamefont {Dirac}(1931)}]{Dirac:1931}%
  \BibitemOpen
  \bibfield  {author} {\bibinfo {author} {\bibfnamefont {P.~A.~M.}\ \bibnamefont {Dirac}},\ }\bibfield  {title} {\bibinfo {title} {Quantised singularities in the electromagnetic field},\ }\href {https://doi.org/10.1098/rspa.1931.0130} {\bibfield  {journal} {\bibinfo  {journal} {Proceedings of the Royal Society of London. Series A, Containing Papers of a Mathematical and Physical Character}\ }\textbf {\bibinfo {volume} {133}},\ \bibinfo {pages} {60} (\bibinfo {year} {1931})}\BibitemShut {NoStop}%
\bibitem [{\citenamefont {Wu}\ and\ \citenamefont {Yang}(1976)}]{Wu:1976}%
  \BibitemOpen
  \bibfield  {author} {\bibinfo {author} {\bibfnamefont {T.~T.}\ \bibnamefont {Wu}}\ and\ \bibinfo {author} {\bibfnamefont {C.~N.}\ \bibnamefont {Yang}},\ }\bibfield  {title} {\bibinfo {title} {Dirac monopole without strings: monopole harmonics},\ }\href {https://doi.org/10.1016/0550-3213(76)90143-7} {\bibfield  {journal} {\bibinfo  {journal} {Nuclear Physics B}\ }\textbf {\bibinfo {volume} {107}},\ \bibinfo {pages} {365} (\bibinfo {year} {1976})}\BibitemShut {NoStop}%
\bibitem [{\citenamefont {Wu}\ and\ \citenamefont {Yang}(1977)}]{Wu:1977}%
  \BibitemOpen
  \bibfield  {author} {\bibinfo {author} {\bibfnamefont {T.~T.}\ \bibnamefont {Wu}}\ and\ \bibinfo {author} {\bibfnamefont {C.~N.}\ \bibnamefont {Yang}},\ }\bibfield  {title} {\bibinfo {title} {Some properties of monopole harmonics},\ }\href {https://doi.org/10.1103/PhysRevD.16.1018} {\bibfield  {journal} {\bibinfo  {journal} {Physical Review D}\ }\textbf {\bibinfo {volume} {16}},\ \bibinfo {pages} {1018} (\bibinfo {year} {1977})}\BibitemShut {NoStop}%
\bibitem [{\citenamefont {Jain}\ and\ \citenamefont {Kamilla}(1997{\natexlab{a}})}]{Jain_Kamilla:1997}%
  \BibitemOpen
  \bibfield  {author} {\bibinfo {author} {\bibfnamefont {J.}~\bibnamefont {Jain}}\ and\ \bibinfo {author} {\bibfnamefont {R.}~\bibnamefont {Kamilla}},\ }\bibfield  {title} {\bibinfo {title} {Composite fermions in the hilbert space of the lowest electronic landau level},\ }\href {https://doi.org/10.1142/S0217979297001301} {\bibfield  {journal} {\bibinfo  {journal} {International Journal of Modern Physics B}\ }\textbf {\bibinfo {volume} {11}},\ \bibinfo {pages} {2621} (\bibinfo {year} {1997}{\natexlab{a}})}\BibitemShut {NoStop}%
\bibitem [{\citenamefont {Wu}\ and\ \citenamefont {Jain}(2013)}]{Wu:2013}%
  \BibitemOpen
  \bibfield  {author} {\bibinfo {author} {\bibfnamefont {Y.-H.}\ \bibnamefont {Wu}}\ and\ \bibinfo {author} {\bibfnamefont {J.~K.}\ \bibnamefont {Jain}},\ }\bibfield  {title} {\bibinfo {title} {Quantum hall effect of two-component bosons at fractional and integral fillings},\ }\href {https://doi.org/10.1103/PhysRevB.87.245123} {\bibfield  {journal} {\bibinfo  {journal} {Physical Review B—Condensed Matter and Materials Physics}\ }\textbf {\bibinfo {volume} {87}},\ \bibinfo {pages} {245123} (\bibinfo {year} {2013})}\BibitemShut {NoStop}%
\bibitem [{\citenamefont {Jain}\ and\ \citenamefont {Kamilla}(1997{\natexlab{b}})}]{Jain:1997}%
  \BibitemOpen
  \bibfield  {author} {\bibinfo {author} {\bibfnamefont {J.}~\bibnamefont {Jain}}\ and\ \bibinfo {author} {\bibfnamefont {R.}~\bibnamefont {Kamilla}},\ }\bibfield  {title} {\bibinfo {title} {Quantitative study of large composite-fermion systems},\ }\href {https://doi.org/10.1103/PhysRevB.55.R4895} {\bibfield  {journal} {\bibinfo  {journal} {Physical Review B}\ }\textbf {\bibinfo {volume} {55}},\ \bibinfo {pages} {R4895} (\bibinfo {year} {1997}{\natexlab{b}})}\BibitemShut {NoStop}%
\bibitem [{\citenamefont {Majumder}\ and\ \citenamefont {Mandal}(2014)}]{Majumder:2014}%
  \BibitemOpen
  \bibfield  {author} {\bibinfo {author} {\bibfnamefont {D.}~\bibnamefont {Majumder}}\ and\ \bibinfo {author} {\bibfnamefont {S.~S.}\ \bibnamefont {Mandal}},\ }\bibfield  {title} {\bibinfo {title} {Neutral collective modes in spin-polarized fractional quantum hall states at filling factors 1/3, 2/5, 3/7, and 4/9},\ }\href {https://doi.org/10.1103/PhysRevB.90.155310} {\bibfield  {journal} {\bibinfo  {journal} {Physical Review B}\ }\textbf {\bibinfo {volume} {90}},\ \bibinfo {pages} {155310} (\bibinfo {year} {2014})}\BibitemShut {NoStop}%
\bibitem [{\citenamefont {Das}\ \emph {et~al.}(2017)\citenamefont {Das}, \citenamefont {Indra},\ and\ \citenamefont {Majumder}}]{Das:2017}%
  \BibitemOpen
  \bibfield  {author} {\bibinfo {author} {\bibfnamefont {D.}~\bibnamefont {Das}}, \bibinfo {author} {\bibfnamefont {M.}~\bibnamefont {Indra}},\ and\ \bibinfo {author} {\bibfnamefont {D.}~\bibnamefont {Majumder}},\ }\bibfield  {title} {\bibinfo {title} {Neutral collective excitation in fractional quantum hall effect at jain series},\ }\href {https://doi.org/10.1016/j.ssc.2017.05.011} {\bibfield  {journal} {\bibinfo  {journal} {Solid State Communications}\ }\textbf {\bibinfo {volume} {260}},\ \bibinfo {pages} {19} (\bibinfo {year} {2017})}\BibitemShut {NoStop}%
\bibitem [{\citenamefont {Viefers}(2008)}]{Viefers:2008}%
  \BibitemOpen
  \bibfield  {author} {\bibinfo {author} {\bibfnamefont {S.}~\bibnamefont {Viefers}},\ }\bibfield  {title} {\bibinfo {title} {Quantum hall physics in rotating bose--einstein condensates},\ }\href {https://doi.org/10.1088/0953-8984/20/12/123202} {\bibfield  {journal} {\bibinfo  {journal} {Journal of Physics: Condensed Matter}\ }\textbf {\bibinfo {volume} {20}},\ \bibinfo {pages} {123202} (\bibinfo {year} {2008})}\BibitemShut {NoStop}%
\bibitem [{\citenamefont {Mandal}\ and\ \citenamefont {Jain}(2002)}]{Mandal:2002}%
  \BibitemOpen
  \bibfield  {author} {\bibinfo {author} {\bibfnamefont {S.~S.}\ \bibnamefont {Mandal}}\ and\ \bibinfo {author} {\bibfnamefont {J.~K.}\ \bibnamefont {Jain}},\ }\bibfield  {title} {\bibinfo {title} {Theoretical search for the nested quantum hall effect of composite fermions},\ }\href {https://doi.org/10.1103/PhysRevB.66.155302} {\bibfield  {journal} {\bibinfo  {journal} {Physical Review B}\ }\textbf {\bibinfo {volume} {66}},\ \bibinfo {pages} {155302} (\bibinfo {year} {2002})}\BibitemShut {NoStop}%
\bibitem [{\citenamefont {Pinczuk}(1996)}]{Pinczuk:1996}%
  \BibitemOpen
  \bibfield  {author} {\bibinfo {author} {\bibfnamefont {A.}~\bibnamefont {Pinczuk}},\ }\bibfield  {title} {\bibinfo {title} {Resonant inelastic light scattering from quantum hall systems},\ }\href {https://doi.org/10.1002/9783527617258} {\bibfield  {journal} {\bibinfo  {journal} {Perspectives in Quantum Hall Effects: Novel Quantum Liquids in Low-Dimensional Semiconductor Structures}\ ,\ \bibinfo {pages} {307}} (\bibinfo {year} {1996})}\BibitemShut {NoStop}%
\end{thebibliography}%
\end{document}